\documentclass[prd, onecolumn, nofootinbib, floatfix]{revtex4} 

\usepackage{amsmath}
\usepackage{graphicx}
\usepackage{dcolumn}
\usepackage{bm}
\usepackage{epsfig}
\usepackage{amssymb,latexsym,mathrsfs}
\usepackage{graphicx}
\usepackage{color}
\usepackage{hyperref}

\hypersetup{
    colorlinks=true,
    linkcolor=red,
    citecolor=blue,
} 

\newcommand{\be}{\begin{equation}}
\newcommand{\ee}{\end{equation}}
\newcommand{\bs}{\begin{split}} 
\newcommand{\bea}{\begin{eqnarray}}
\newcommand{\eea}{\end{eqnarray}}

\newcommand{\ode}{\Omega_{\rm de}}
\newcommand{\oode}{{\mathcal O}(\Omega_{\rm de})}

\newcommand{\geff}{G_{\rm eff}} 
\newcommand{\geffh}{G_{\rm eff}^\Phi} 
\newcommand{\geffs}{G_{\rm eff}^\Psi} 
\newcommand{\geffsh}{G_{\rm eff}^{\Psi+\Phi}}

\def\nn{\nonumber} 

\begin{document}

\title{Is the Effective Field Theory of Dark Energy Effective?} 
\author{Eric V.\ Linder$^1$, Gizem Seng{\"o}r$^2$, Scott Watson$^2$} 
\affiliation{$^1$ Berkeley Center for Cosmological Physics \& Berkeley Lab, 
University of California, Berkeley, CA 94720, USA\\ 
$^2$ Department of Physics, Syracuse University, Syracuse, NY 13244, USA} 

\begin{abstract}
The effective field theory of cosmic acceleration systematizes possible 
contributions to the action, accounting for both dark energy and 
modifications of gravity. Rather than making model dependent assumptions, 
it includes all terms, subject to the required symmetries, with four (seven) 
functions of time for the coefficients. These correspond respectively to 
the Horndeski and general beyond Horndeski class of theories. We address 
the question of whether this general systematization is actually effective, 
i.e.\ useful in revealing the nature of cosmic acceleration when compared 
with cosmological data. The answer is no and yes: {\it there is no simple time 
dependence of the free functions\/} -- assumed forms in the literature are poor fits, 
but one can derive some general characteristics in early and late time 
limits. For example, we prove that the gravitational slip must restore to general 
relativity in the de Sitter limit of Horndeski theories, and why it doesn't more 
generally. 
We also clarify the relation between the tensor and scalar sectors, 
and its important relation to observations; in a real sense the expansion history 
$H(z)$ or dark energy equation of state $w(z)$ is $1/5$ or less of the functional 
information! In addition we discuss the de Sitter, 
Horndeski, and decoupling limits of the theory utilizing Goldstone techniques. 
\end{abstract} 

\date{\today} 

\maketitle

\section{Introduction} 

Cosmic acceleration is fundamental physics beyond the Standard Model, 
with potentially revolutionary implications for the nature of gravitation 
and spacetime, the quantum vacuum, and high energy physics. Its origin 
may lie somewhere along the spectrum of a single number (the cosmological 
constant), a single canonical minimally coupled scalar field (quintessence), 
a scalar-tensor theory, or a multiply coupled, ``braided'' modification 
altering the nature of tensor gravity and allowing superluminal propagation. 
Moreover, there is no clear first principles derivation of any of these, 
pointing to a particular model. Thus the task of revealing the nature of 
cosmic acceleration by comparing models to observations seems gargantuan. 

An alternative is to take a bottom-up approach and construct a Lagrangian 
by writing the most general theory possible which can account for the cosmic 
acceleration, 
where the allowed operators are restricted by a combination of symmetries and
theoretical self-consistency. This is the goal of the Effective 
Field Theory (EFT) approach \cite{EFTrevs}.  EFT comprises two parts: 
1) integrating out the physics of the high energy degrees of 
freedom to derive an effective low energy theory suitable for comparison to 
observations in our low energy world (where low energy is relative, e.g.\ for 
inflation this is low relative to the Planck energy), and 2) including all 
low energy degrees of freedom allowed by the symmetries. When spontaneous symmetry 
breaking is present, as in a time dependent background, 
focusing on the Goldstone bosons of the broken symmetry leads to a 
particularly powerful approach, 
since the requirement to non-linearly realize the broken symmetry 
forces relations between the free parameters of the low energy effective 
theory -- the corresponding operators exhibit universal behavior.

The EFT approach has been successfully used in many circumstances, including 
the early cosmic acceleration of inflation \cite{Cheung:2007st,Senatore:2010wk} and the 
theory of dark matter \cite{Baumann:2010tm,Carrasco:2012cv}. EFT has also been written down for 
late cosmic acceleration, i.e.\ dark energy (EFTDE) \cite{Creminelli:2008wc,Park:2010cw,Gubitosi:2012hu,bloomfield}, but it is not 
clear that its {\it application\/} to dark energy has been successful. 
Three aspects of the EFTDE present challenges that are somewhat distinct from 
the cases above, in both theoretical and practical issues: 1) The coefficients of 
the operators in the action are time dependent functions rather than 
constants; 
2) In the most general EFTDE there are more coefficients than observables; 
3) The primary observational constraints on these coefficients occur 
while the universe is (nearly) matter dominated, whereas the theory is constructed by using the spontaneous breaking of
time translation invariance during (complete) dark energy domination.  Indeed, for constraints on modified gravity in the current epoch
it is known that for long distance (IR) modifications of gravity near the Hubble scale the time 
it takes for such modifications to result in local, observational consequences is longer than the current age of the universe \cite{ArkaniHamed:2003uy}.
Moreover, in many modified gravity theories screening and/or decoupling from gravity lead to a lack of observational consequences on short distance scales. 

We address here aspects of these questions. Basically, can the effective 
field theory of dark energy be effective? We take care to bridge the 
theoretical, phenomenological, and observations descriptions of cosmic 
acceleration and in particular investigate both the scalar and tensor 
aspects of the theory. One of the recent exciting developments in cosmic 
acceleration has been the identification of relations between properties 
of the matter density field (scalar sector) and gravitational wave 
propagation (tensor sector) \cite{bellini,amendola,slip}, offering 
new observational connections. 

In Sec.~\ref{sec:defs} we briefly review the different methods of 
describing the impact of beyond Einstein physics from the theory, 
phenomenology, and observational points of view, giving the translations 
between them. The functions defined in these ways are examined more 
closely in Sec.~\ref{sec:param}, with numerical solutions given for the 
specific case of Galileon gravity, establishing that simple parameterizations 
as often used in the literature to place observational constraints are 
insufficient and biased. We also derive some general relations in the early, 
matter dominated, and late time, asymptotic de Sitter (dS) limits. 
In Sec.~\ref{sec:dS} we focus on the Goldstone mode in the EFTDE.
We use this to examine the dS, Horndeski, and decoupling limits of the theory.
Using these limits and the Goldstone method we are able to reproduce some of the key findings from
the previous sections and we discuss how viable modifications of gravity fall into two primary 
classes based on their decoupling behavior. 
We discuss the implications for the effectiveness 
of EFTDE to explore cosmic acceleration in Sec.~\ref{sec:concl}.

\section{Describing Cosmic Acceleration in Theory and Observations} 
\label{sec:defs} 

EFT provides an action with terms of particular forms determined by the 
allowed symmetries, and coefficients that in the case of dark energy 
(used generically as including modified gravity) are time dependent 
functions. These coefficients have dimensions of mass and so can 
be illustratively written as $M_i(t)$. Constraining the $M_i(t)$ gives 
specific information about the class of theory, since different theories 
predict different terms in the action, i.e.\ some $M_i$ are zero. 
Observational quantities on the other hand deal with measurements of 
density contrasts and particle motions (e.g.\ velocities or gravitational 
lensing). These can be phrased in terms of modified Poisson equations 
and the relation between metric potentials, i.e.\ the effective gravitational 
coupling strength (generalized Newton's constant) and the gravitational slip. 
Phenomenology attempts to sit between the theory and observational 
quantities, e.g.\ dealing with ``property'' functions characterizing and 
following from the 
equations of motion. We briefly review each, and connect the quantities.

\subsection{EFT Action and Functions} \label{sec:action} 

The EFTDE quadratic action in the Jordan frame, following the parameter 
notation of \cite{bloomfield} and working with metric signature 
$(-,+,+,+)$, is 
\bea \label{eftde}
\nn S_2&=&\int d^4x \, \sqrt{-g} \Bigg[\frac{m^2_0}{2}\Omega(t)R-\Lambda(t)-c(t)g^{00}+\frac{M^4_2(t)}{2}(\delta g^{00})^2-\frac{\bar{M}_1^3(t)}{2} \delta K\delta g^{00} \\
&-& \frac{\bar{M}_2^2(t)}{2}  \delta K^2 -\frac{\bar{M}_3^2(t)}{2} \delta K^{\;\; \mu}_\nu\delta K^{\;\;\nu}_{\mu}+\frac{\hat{M}^2(t)}{2} \delta R^{(3)}\delta g^{00} +m^2_2(t) \, \partial_i g^{00} \partial^i g^{00}+ {\cal L}_m\Bigg],
\eea
where $\delta g^{00}=g^{00} +1$ and $\delta R^{(3)}$ are perturbations of the time-component of the metric and the spatial curvature, respectively, 
and given the normal to constant time hypersurfaces
\be
n_\mu = \frac{\partial_\mu t}{\sqrt{-\partial_\sigma t \partial^\sigma t}} = \delta^0_\mu,
\ee
we have the perturbation of the extrinsic curvature
\be
\delta K_{\mu \nu}=K_{\mu \nu} - K_{\mu \nu}^{(0)}=K_{\mu \nu} + 3H\left( g_{\mu \nu} + n_\mu n_\nu \right) \ , 
\ee 
where its trace $\delta K=\delta K_\mu^\mu$, and $H(t)$ is the Hubble expansion 
parameter. 

The action involves background quantities $\Lambda(t)$ and $c(t)$, and linearly 
perturbed quantities involving the mass coefficients. In the simple case of 
quintessence the background terms correspond to the potential 
$\Lambda(t)=V(\phi(t))$ and the kinetic term $c(t) = \dot{\phi}^2/2$ of the 
scalar field.  In more involved models these background expressions can be more 
complicated. 

Note that we wrote the gravitational coupling to the Ricci tensor as 
$m_0^2\Omega(t)$, to show the explicit mass dimension in the constant 
$m_0$, leaving the time dependence in $\Omega(t)$. When $\Omega$ is 
constant (e.g.\ unity) we can identify $m_0^2 \rightarrow m_p^2$ where we 
use reduced Planck mass units $m_p=1/\sqrt{8\pi G_N} = 2.4 \times 10^{18}$ GeV. 
In general, $m_0^2\Omega(t)$ is a single quantity, which we will sometimes 
denote $p$. While $m_0^2\Omega(t)$ is a homogeneous background quantity 
it is important to note that given our bottom-up approach (and without knowledge of the UV completion of the theory) 
there are not enough equations to fix $\Omega(t)$ uniquely from the background evolution \cite{Park:2010cw}.  This 
parameter must be determined by enforcing symmetries and constraints coming 
from considering the perturbations which start at quadratic order in the 
action. 

The background equations of motion (which are the tadpole cancellation 
conditions, discussed below) in the Jordan frame are given by 

\bea 
3 H^2+3\frac{\dot{\Omega}(t)}{\Omega(t)} H &=& \frac{1}{m_0^2\Omega(t)} \left( \
c+\Lambda + \rho_m \right) \label{eq:friedh}\\
2 \dot{H} +\frac{\ddot{\Omega}(t)}{\Omega(t)} +\frac{\dot{\Omega}(t)}{\Omega(t)\
} H &=& - \frac{1}{m_0^2\Omega(t)} \left( 2 c +\rho_m + p_m\right) \ .  \label{eq:friedh2}
\eea 

Several comments are in order. For arbitrary $\Omega(t)$ we have a 
scalar-tensor theory (which in some sense is {\it not\/} a true modification 
of gravity as it does not imply a change in the propagator of the spin-2 
graviton) and the Jordan and Einstein frames are manifestly different 
although they will lead to the same observables.  We use these background 
equations to eliminate tadpole terms from the action.  In other words, 
when we expand the first (gravitational) term in Eq.~(\ref{eftde}) 
we want the perturbations from that term to be canceled at the linear 
level by the terms coming from $c(t)\delta g^{00}$ and $\Lambda(t)$.  
This is equivalent to solving the background equations for $c$ and 
$\Lambda$ and plugging them back into the action. 
After this procedure the action that results from Eq.~(\ref{eftde}) 
is the most general theory possible for the scalar perturbations (at 
the quadratic level) about a spatially homogeneous background 
that spontaneously breaks time diffeomorphism invariance 
\cite{Gubitosi:2012hu,bloomfield}.

There are seven free parameters that describe the most general 
theory \cite{Gubitosi:2012hu,bloomfield}: 
\be \label{params}
\left[ m_0^2\Omega(t), \bar M_1^3(t), M_2(t), \bar M_2(t), \bar M_3(t), \hat M(t),
m_2(t) \right] .
\ee 
Given these parameters it is possible to characterize all existing models -- 
see Table~\ref{tab:models} in Appendix~\ref{sec:operator}.
For example, linearized Horndeski theory, or equivalently generalized Galileon theory, 
is reproduced by the parameter choices 
\be \label{hornparams}
m_2=0;\quad 2 \hat{M}^2=\bar{M}_2^2=-\bar{M}_3^2 \qquad{\rm [Horndeski]} 
\ . 
\ee 
This class of theories has received notable attention because the resulting equations of motion are second order in time and space derivatives, which ensures they do not suffer from
an Ostrogradsky instability and there is a well defined Cauchy problem. 

The additional parameters in Eq.~\eqref{params} however allow for the 
construction of more general models, raising the question -- are these 
theories sound and stable? We explain a simple rationale for their validity 
in Appendix~\ref{apx:beyond}. The upshot is that by construction the EFTDE approach renders these theories stable within their regime of applicability regardless 
of the number of time and/or space derivatives. We emphasize that this does  
{\em not\/} imply the more restricted Horndeski-like theories are uninteresting. Instead, we are pointing out that the EFTDE approach provides a complete framework for establishing which proposals for dark energy are observationally interesting when confronted with existing data.  If a class of models is established as favored by the data it will then be an important endeavor to establish the high energy completion of such a model. Moreover, Horndeski-like theories provide an important class of models where a non-perturbative formulation seems possible.  It is an important question whether such models can be embedded in a complete theory of quantum gravity, but we see that in the observationally relevant regime 
these theories are captured within the EFTDE through the parameter selection given in Eq.~\eqref{hornparams}. 
Thus, if observations can rule out such models in the EFT regime this would suggest model building should go in a different direction.  

We would also like to emphasize that the seven parameters of Eq.~\eqref{params} correspond to the most general action at {\em quadratic\/} level in 
the perturbations. 
Going beyond the quadratic level leads to additional parameters that need to be determined through a combination of theoretical and observational constraints.
For investigations during the linear regime of structure formation (where perturbations are small) these seven parameters should provide the leading contributions, but even then we will see below that 
there are only four observational quantities to restrict the seven free parameters.  That is, there is an {\em inverse\/} problem. 
This is a familiar situation from particle physics, where collider data places constraints on those EFT parameters, but many different proposals for physics beyond the standard model 
can lead to the same observational predictions. 
We have a similar challenge to address here with degeneracies in the map 
between the EFT parameters and experimental observations. 
The more general (beyond Horndeski) EFTDE approach most likely will require 
novel observations to determine fully the nature of dark energy.

\subsection{Observational Functions} \label{sec:obs} 

The EFTDE approach allows us to specify the expansion history of 
the background universe, which is described by the Hubble parameter $H(t)$.
The remaining parameters are also connected to direct observables, e.g. the 
growth and motion of structure (perturbations). The impact of the theory on 
the perturbed quantities is conveniently described in terms of the modified 
Poisson equations for non-relativistic and relativistic particles, e.g.\ 
galaxies and light. 

Working in the conformal Newtonian gauge, the perturbed metric is 
\be
ds^2 = -\left( 1 + 2 \Psi \right) dt^2 + a^2(t) \left( 1 - 2 \Phi \right) 
d\vec{x}^2 \ , 
\ee
where we assume vanishing spatial curvature. The modified Poisson equations 
can then be written as 
\bea 
& &   \nabla^{2} \Psi = 4\pi a^{2} G^{\Psi}_{\rm eff} \rho_m\,\delta_ m \\  
& &   \nabla^{2} (\Psi+\Phi) = 8\pi a^{2} G^{\Psi+\Phi}_{\rm eff} 
\rho_m\,\delta_m \,, 
\eea 
where the (time and scale dependent) $\geff$ are modified Newton's constants. 
In the quasi-static 
regime they can be viewed as giving the gravitational coupling strength. 
The quantity $\rho_m$ is the matter density, and 
$\delta_m=\delta\rho_m/\rho_m$ is the density perturbation. 
The first equation is central to the motion of non-relativistic 
matter and hence the growth of massive structures, while the second governs 
the motion of relativistic particles, and hence the propagation 
of light \cite{bert11}. 

While $\geffs$ and $\geffsh$ are tied directly to observables 
through these equations, it is also common in the literature to use 
$\geffs$ and a gravitational slip defined as 
\be 
\eta=\frac{\Psi}{\Phi}=\frac{\geffs}{2\geffsh-\geffs} \ . 
\ee  
In \cite{bellini} they use 
\be 
\bar\eta=\frac{2\Psi}{\Psi+\Phi}=\frac{2\eta}{1+\eta}=\frac{\geffs}{\geffsh} 
\ , 
\ee 
which is somewhat more closely tied to observations, and we will use this 
version of the slip. 

In addition, observation of gravitational waves (e.g.\ through cosmic 
microwave background B-mode polarization, pulsar timing arrays, or directly 
with gravitational wave antennas) would allow the tensor 
sector to be probed. The wave propagation equation involves three quantities: 
the propagation speed $c_T(t)$, the friction term, and modifications to the 
source term. The last one arises from anisotropic stress, which also 
sources the gravitational slip, so we basically have two more 
``observable'' functions from the tensor sector. This point has been 
somewhat underappreciated in the literature -- the scalar sector gives 
two functions worth of information on the theory and the tensor sector 
gives two functions worth as well. Together they add up to four functions, 
equivalent to that of the theory description in the Horndeski framework. 
In beyond Horndeski theories, the gravitational wave source term gets 
additional contributions and there can be a mass term as well. 
As mentioned in the previous section the Horndeski correspondence between 
the observables and the free parameters is the result of working with a 
special class of models, at the quadratic level.

\subsection{Property Functions} \label{sec:alpha} 

In \cite{bellini}, they develop ``property'' functions that describe 
classes of theories phenomenologically, characterizing properties 
such as the kineticity -- 
kinetic structure of the scalar field, affecting the sound speed, 
braiding -- mixing of the scalar kinetic terms with the metric, affecting 
dark energy clustering, running Planck mass -- leading to anisotropic 
stress, and tensor wave speed. These give four time dependent functions 
$\alpha_K(t)$, $\alpha_B(t)$, $\alpha_M(t)$, and $\alpha_T(t)$ that 
enter in the equations of motion. 

The property functions play a role midway between the directly theoretical 
exhibition of the action functions and the ties to actual data of the 
observational 
functions. Just as one might approach quintessence through the more closely 
theoretical potential $V(\phi)$ or the more phenomenological equation of 
state $w(t)$, which type of function is used is mostly a matter of taste. 
In the next section we give translations between these functions.

\subsection{Translating between Functions} \label{sec:trans} 

We assemble here the key equations relating the action, property, 
and observational functions to each other. 

The $\alpha_i$'s from Table~2 of \cite{bellini} for the action 
Eq.~\eqref{eftde} above are 
\bea
\alpha_M&=&\frac{\dot \Omega+\frac{\dot{\bar{M}}_2^2}{m^2_0}}{H \Omega+
H\frac{\bar{M}_2^2}{m^2_0}} =\frac{(\dot p/p)(1+N)+\dot N}{H(1+N)}\\
\alpha_K&=&\frac{2c+4M^4_2}{m^{2}_0\left(H^2 \Omega+H^2\frac{\bar{M}_2^2}{m^2_0}\right)} =2\frac{c+2M_2^4}{H^2 p(1+N)}\\
\alpha_B&=&-\frac{\bar{M}_1^3+m^2_0\dot{\Omega}}{m^{2}_0\left(H \Omega+
H\frac{\bar{M}_2^2}{m^2_0}\right)} =-\frac{\dot p+\bar M_1^3}{Hp(1+N)}\\
\alpha_T&=&-\frac{\bar{M}_2^2}{m^{2}_0\left(\Omega+\frac{\bar{M}_2^2}{m^2_0}\right)} =-\frac{N}{1+N} \ , 
\eea 
where we define $p(t)=m_0^2\Omega(t)$ and $N(t)=\bar M_2^2/p$. 

The converse transformation is 
\bea 
N&=&\frac{-\alpha_T}{1+\alpha_T} \qquad;\qquad 1+N=(1+\alpha_T)^{-1} 
=c_T^{-2} \label{eq:nalpha}\\ 
\frac{\dot p}{p}&=&H\alpha_M-\frac{\dot N}{1+N}\\ 
\bar M_1^3&=&-Hp\frac{\alpha_B}{1+\alpha_T}-\dot p=-Hp\left(\frac{\alpha_B}{1+\alpha_T}+\alpha_M-\frac{N'}{1+N}\right)\\ 
c+2M_2^4&=&\frac{1}{2}\,\alpha_K H^2p(1+N) \ , \label{eq:malpha} 
\eea 
where prime denotes $d/d\ln a$. 

We see that $N$ depends only on $\alpha_T$, $p$ depends only on $\alpha_M$ 
and $\alpha_T$ (and the background expansion $H$, which we take as given), 
$\bar M_1^3$ depends on $\alpha_B$, $\alpha_M$, and 
$\alpha_T$, and $\alpha_K$ only enters for the combination $c+2M_2^4$. 
Since the observables or their proxies, e.g.\ the gravitational coupling 
$G_{\rm eff}$ and gravitational slip $\bar\eta$ entering 
the growth of structure only depends on the $\alpha_i$, this seems 
to imply that $c$ and $M_2^4$ cannot be separately determined, only the 
particular combination $c+2M_2^4$. Some quantities that are not 
directly observable, such as the scalar sound speed $c_s$ and the quantity $C_3$ 
below, do depend on $c$ separately. Beyond the quasistatic approximation (e.g.\ 
near horizon scales), $M_2$ may appear separately \cite{13046712}. 

The observational functions are related to the theory and phenomenology 
functions by the expressions below. 
Following \cite{bloomfield} and working in the Newtonian limit we have
\be
4 \pi G_{\rm{eff}}=\frac{C_3-C_1 B_3}{A_1(B_3C_2-B_1C_3)+A_2(B_1C_1-C_2)+A_3(C_3-C_1B_3)} \ , \label{eq:geffgen} 
\ee 
where the derivation and the expressions for the $A_i$, $B_i$, and $C_i$ 
are given in Appendix~\ref{sec:apxabc}. 

In the particular case of Horndeski theory and the dS limit 
we find 
\be
4 \pi G_{\rm{eff}}=
\frac{1}{2m_0^2\Omega}\left( {1+\frac{\bar{M}_2^2}{m_0^2\Omega}+\frac{\bar{M}_1^3}{2m_0^2\Omega H}} \right)^{-1} \qquad [{\rm Horndeski,\ de\ Sitter\ limit}] \ . 
\ee 
If instead of Horndeski and the dS limit one focuses only on the 
background quantities (all $\bar{M}_i=0$, and $m_2=0=\hat M$) we have 
\be \label{eq:geffbgd} 
4 \pi G_{\rm{eff}}=\frac{1}{2m_0^2 \Omega}\left(1+\frac{(\dot{\Omega}/\Omega)^2}{4c/(m_0^2\Omega)+3(\dot{\Omega}/\Omega)^2} \right) \qquad [{\rm background}]\ . 
\ee 
This corresponds to Brans-Dicke type modifications of gravity.
In either case we recover $4\pi\geff=1/(2m_p^2)$ for the Einstein-Hilbert 
action. 

For the gravitational slip, 
in the Newtonian limit the general expression is 
\be 
\bar{\eta}=\frac{2(C_3-C_1 B_3)}{B_3 C_2-B_1 C_3+C_3-C_1 B_3} \ . 
\label{eq:etagen} 
\ee 
Of great interest is that in the dS limit of Horndeski theory 
one immediately finds that 
\be  \label{eq25}
\bar{\eta} \rightarrow 1 \qquad [{\rm Horndeski,\ de\ Sitter\ limit}] \ . 
\ee 
This derivation of a general property for an entire family of theories 
is a major success for EFTDE. 
This property had been derived for the restricted case of covariant Galileon 
theory in \cite{slip}, where it was shown that despite this vanishing of 
deviations of the gravitation slip, 
the tensor sector still showed clear departures from general relativity. 

Apart from Horndeski theory and the dS limit, if one just turns on 
the background operators then working in the Newtonian limit 
gives 
\be \label{eq:etabgd} 
\bar{\eta}=\frac{c/(m_0^2\Omega)+(\dot{\Omega}/\Omega)^2}{c/(m_0^2\Omega) + (3/4)(\dot{\Omega}/\Omega)^2} \qquad[{\rm background}]\ . 
\ee 

As far as the tensor wave speed, that has the simple general expression 
\be 
c_T^2 =  \left({1-\frac{\bar{M}_3^2}{m_0^2 \Omega}}\right)^{-1}\ . 
\label{eq:ctgen} 
\ee 
Note that the graviton is only affected by $\bar{M}_3$; in the Horndeski case 
$\bar{M}_3=-\bar{M}_2$ and so
\be 
c_T^2 = (1+N)^{-1}= \left({1+\frac{\bar{M}_2^2}{m_0^2 \Omega}}\right)^{-1} \qquad [{\rm Horndeski}] \ . 
\ee 
Here $\Omega(t)$ 
only enters because it defines the effective Planck mass -- no time 
derivatives of $\Omega(t)$ appear (though they do, in terms of $\alpha_M$, 
i.e.\ the running of the Planck mass, in the gravitational wave propagation 
friction term). 
In the absence of the extrinsic curvature terms, $\bar{M}_3 \rightarrow 0$ 
and we recover $c_T=1$. (See Appendix~\ref{sec:apxmod} for a discussion of 
the meaning of modification of gravity.) 

Finally, the observable functions are related to the property functions 
by \cite{bellini} 
\be 
\frac{\geffh}{G_N}= \frac{2m_p^2}{M_\star^2} 
\frac{[\alpha_B(1+\alpha_T)+2(\alpha_M-\alpha_T)]+\alpha_B'}{(2-\alpha_B)[\alpha_B(1+\alpha_T)+2(\alpha_M-\alpha_T)]+2\alpha_B'} \ , \label{eq:geff}
\ee 
where $M_\star^2=m_0^2\Omega+\bar M_2^2=m_0^2\Omega/(1+\alpha_T)$; 
note $\alpha_M=(\ln M_\star^2)'$. 

The gravitational slip $\bar\eta$ is given by 
\be
\bar\eta-1=
\frac{(2+2\alpha_M)[\alpha_B(1+\alpha_T)+2(\alpha_M-\alpha_T)]+(2+2\alpha_T)\alpha_B'}{(2+\alpha_M)[\alpha_B(1+\alpha_T)+2(\alpha_M-\alpha_T)]+(2+\alpha_T)\alpha_B'} \ , \label{eq:etafull} 
\ee 
and the tensor wave speed is 
\be 
c_T^2=1+\alpha_T \ . 
\ee 

Whether using the EFT action functions, property functions, 
or observational functions, one always has another function $H(t)$ 
describing the background. In addition, the tensor wave speed $c_T(t)$ 
is evident in all three, appearing as itself in the observational, 
as $\alpha_T=c_T^2-1$ in the phenomenological, and essentially as $\bar M_3$ 
in the theory approaches.

\section{Functions to Parameters?} \label{sec:param} 

Since EFT provides four or more functions of time in the $M_i$ or $\alpha_i$, 
apart from 
the background expansion $H(t)$, constraints from observations will be 
difficult in full generality. Usually one needs to compress the number of 
degrees of 
freedom by a lower dimensional parametrization. Attempts to do so by choosing 
power law dependence on scale factor $a$, or proportionality to the 
contribution of the effective dark energy density $\ode$ 
(usually assuming a constant 
energy density, i.e.\ a cosmological constant) 
to the total dynamics as a function of time have appeared in the literature. 
We find below that even for restricted classes of theories these are 
vastly oversimplified and hence inaccurate.

\subsection{General Expressions and Numerical Results} \label{sec:genl} 

The property functions $\alpha_i$ are built out of ratios of several 
Lagrangian terms and there is no obvious reason why they should have a 
simple time dependence, especially when terms are of comparable magnitude 
and can interact and cancel as they evolve. The one exception might be 
at high redshift when the deviations from General Relativity (GR) are small; 
there one Lagrangian term may initially dominate, and hence also drive 
the effective dark energy density. This should not generically hold at 
late times however when cosmic acceleration becomes important -- and the vast 
majority of observations enter. 

We illustrate this for the case of covariant Galileons. The Galileon 
subclass of the Horndeski Lagrangian involves four terms, with constant 
coefficients $c_2$, 
$c_3$, $c_4$, $c_5$ all presumably of order unity. These combine into 
functions of time $\kappa_i$ which can be translated into the property 
functions $\alpha_i$ (see the Appendix of \cite{slip}). For example, 
\bea 
\alpha_B&=&\frac{2\kappa_5\,x}{\kappa_4}\\ 
&=&\frac{4c_3\bar H^2x^3-24c_4\bar H^4 x^4+30c_5 \bar H^6 x^5+8c_G \bar H^2 
x^2}{-2+3c_4 \bar H^4 x^4-6c_5 \bar H^6 x^5-2c_G \bar H^2 x^2} \,, 
\eea 
where $x=(1/m_p)d\phi/d\ln a$ and $\bar H=H(a)/H_0$. 

At early times, $\bar H$ is large and the $c_5$ term may be expected to 
dominate in both the effective dark energy density and the property 
functions, i.e.\ $\alpha_i\propto c_5\propto \Omega_{\rm de}$. But this 
{\it only\/} holds when the dark energy density is small. When it 
begins to contribute significantly to the dynamics, eventually leading 
to cosmic acceleration, then $\bar H\sim{\mathcal O}(1)$ and all terms 
become of the same order -- but with the possibility of opposite signs 
which can 
lead to cancellations in the denominator and large swings in the values 
of the $\kappa_i$ or $\alpha_i$. 

\begin{figure}[htbp!]
\includegraphics[width=0.48\columnwidth]{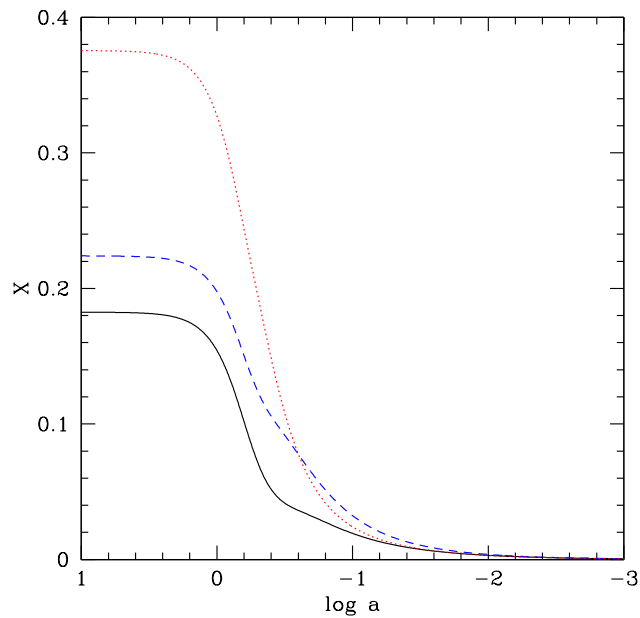} 
\caption{
The canonical kinetic term $X=(1/2)\dot\phi^2/m_p^2$ is plotted vs 
the log of the scale factor $a$ for the Galileon gravity cases 
corresponding to Fig.~1 (solid black), 2 (dashed blue), and 3 (dotted red) 
respectively of \cite{slip}. Note that $X$ cannot be easily fit by a low 
order polynomial or simple function. 
} 
\label{fig:kin} 
\end{figure}

As one numerical example of the full evolution, consider the kinetic 
function $X=\dot\phi^2/2=\bar H^2 x^2/2$. This not only enters the 
Galileon Lagrangian terms and the $\alpha_i$ as above, but more generally 
the Horndeski Lagrangian involves functions of $X$. Figure~\ref{fig:kin} 
shows that the functional dependence of $X(a)$ itself is unlikely to 
be easily parametrized by a power law scale factor dependence or other 
simple form.

In fact, the situation is even more complicated in that the functions 
that connect directly to observables, such as the effective gravitational 
coupling in the modified Poisson equations or the gravitational slip 
between the metric potentials, are themselves ratios of products of the 
$\alpha_i$. This is evident from Eqs.~(\ref{eq:geff})-(\ref{eq:etafull}). 

Essentially, the observables constrain functions (e.g.\ $G_{\rm eff}$) that 
are a ratio of sums of products (of $\alpha_i$) that are a ratio of sums 
(of $c_i$). Illustratively, 
\be 
{\rm Observable}\sim\frac{\sum\prod\left(\frac{\sum f_1(H,X)}{\sum f_2(H,X)}\right)}{\sum\prod\left(\frac{\sum f_3(H,X)}{\sum f_4(H,X)}\right)} \ . 
\ee 
There is no reason to expect a simple time dependence should provide an 
accurate parametrization. 

\begin{figure}[htbp!]
\includegraphics[width=0.48\columnwidth]{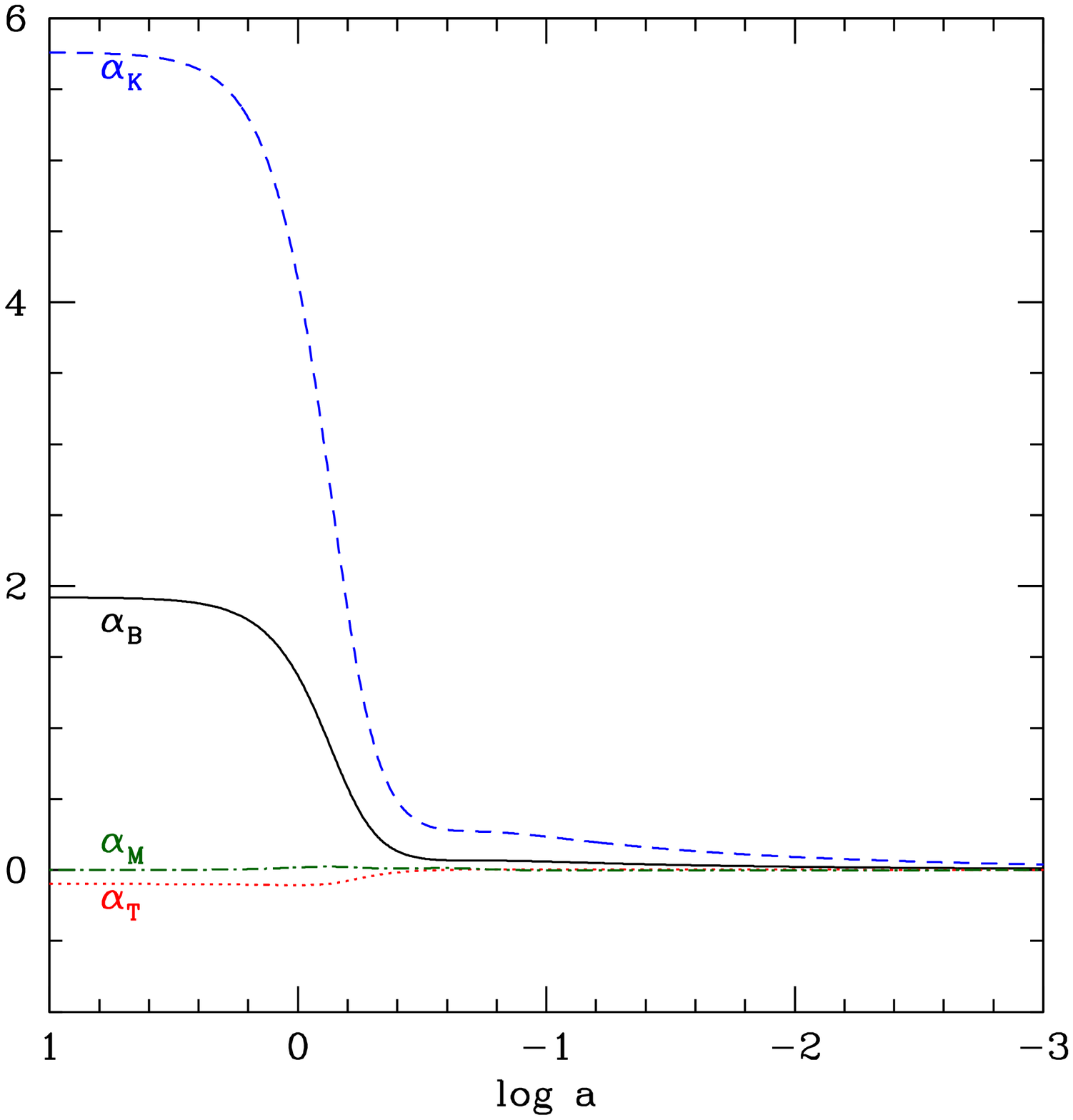} 
\includegraphics[width=0.48\columnwidth]{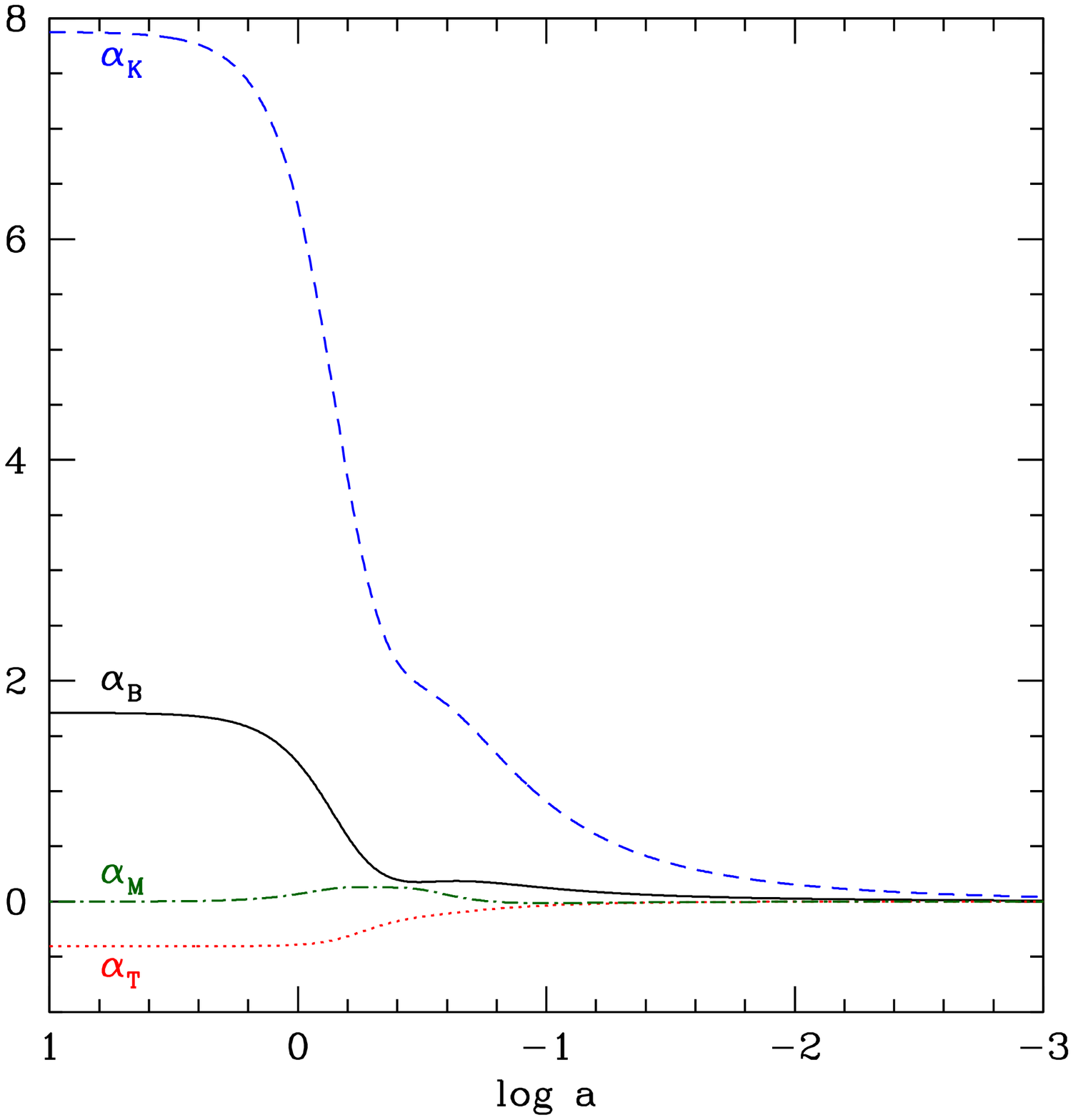}\\ 
\includegraphics[width=0.48\columnwidth]{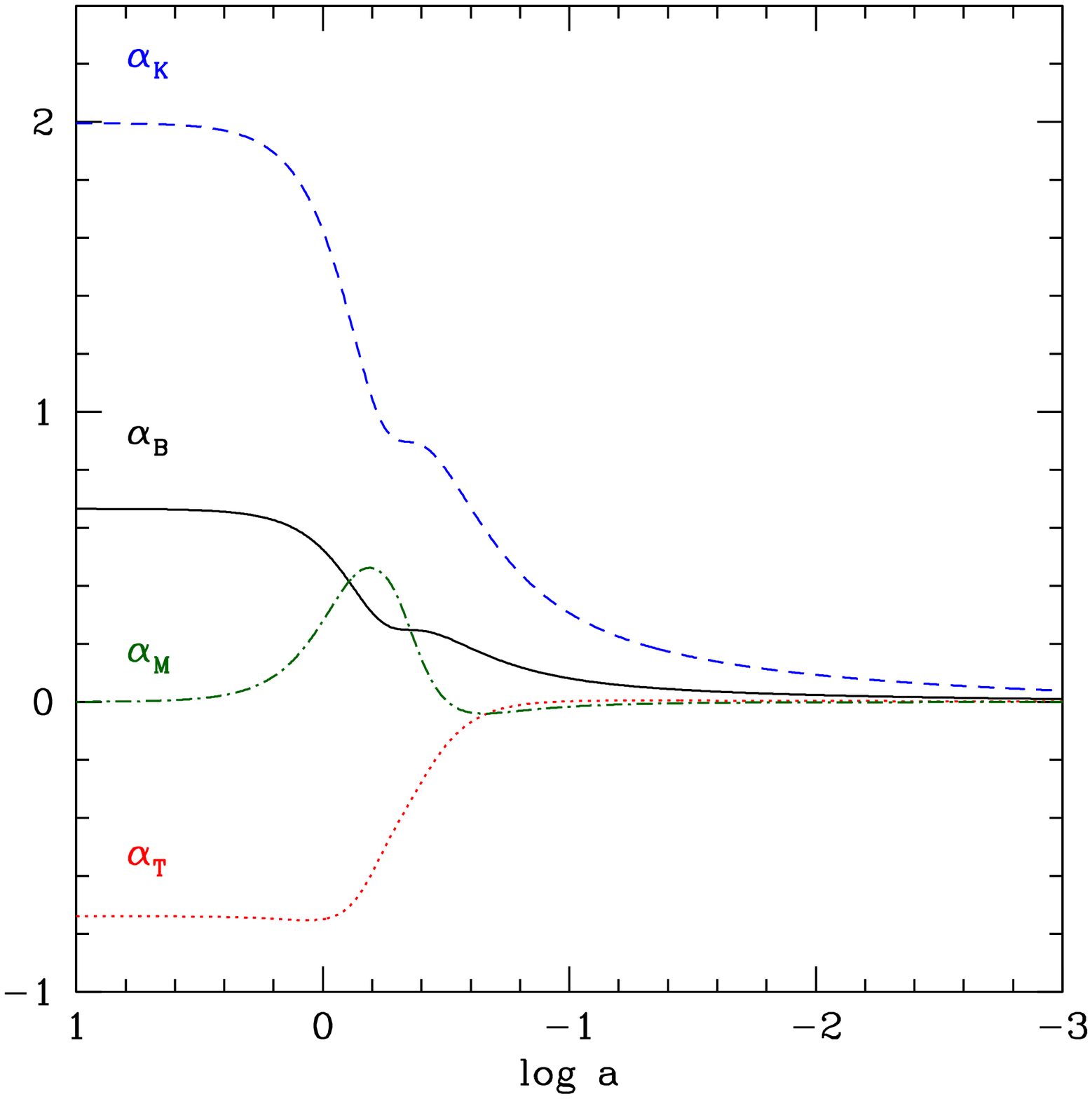} 
\caption{
The property functions $\alpha_i$ are plotted vs the logarithm of the 
scale factor for the Galileon gravity cases corresponding to Figs.~1, 2, 
and 3 respectively of 
\cite{slip}, showing nonmonotonic and complicated dependence of these 
functions with scale factor. 
} 
\label{fig:alfa} 
\end{figure}

Figure~\ref{fig:alfa} shows the 
numerical solutions for the $\alpha_i(a)$ for several Galileon cases.

Again, the only simple time dependence comes in the early time limit 
where indeed all quantities are proportional to $\Omega_{de}$, as 
derived for Galileon gravity in Eqs.~57-61 of \cite{gal1112} and for 
EFT in the next subsection (but deriving the proportionality 
constant requires solving the equations of motion, and so is difficult to 
do in a model independent manner), and in the late time dS 
limit where all quantities becomes time independent. During redshifts 
$z\approx 0-10$, 
where almost all observations lie with constraining power on modified 
gravity causing cosmic acceleration, simple time dependences fail. 

\begin{figure}[htbp!]
\includegraphics[width=0.48\columnwidth]{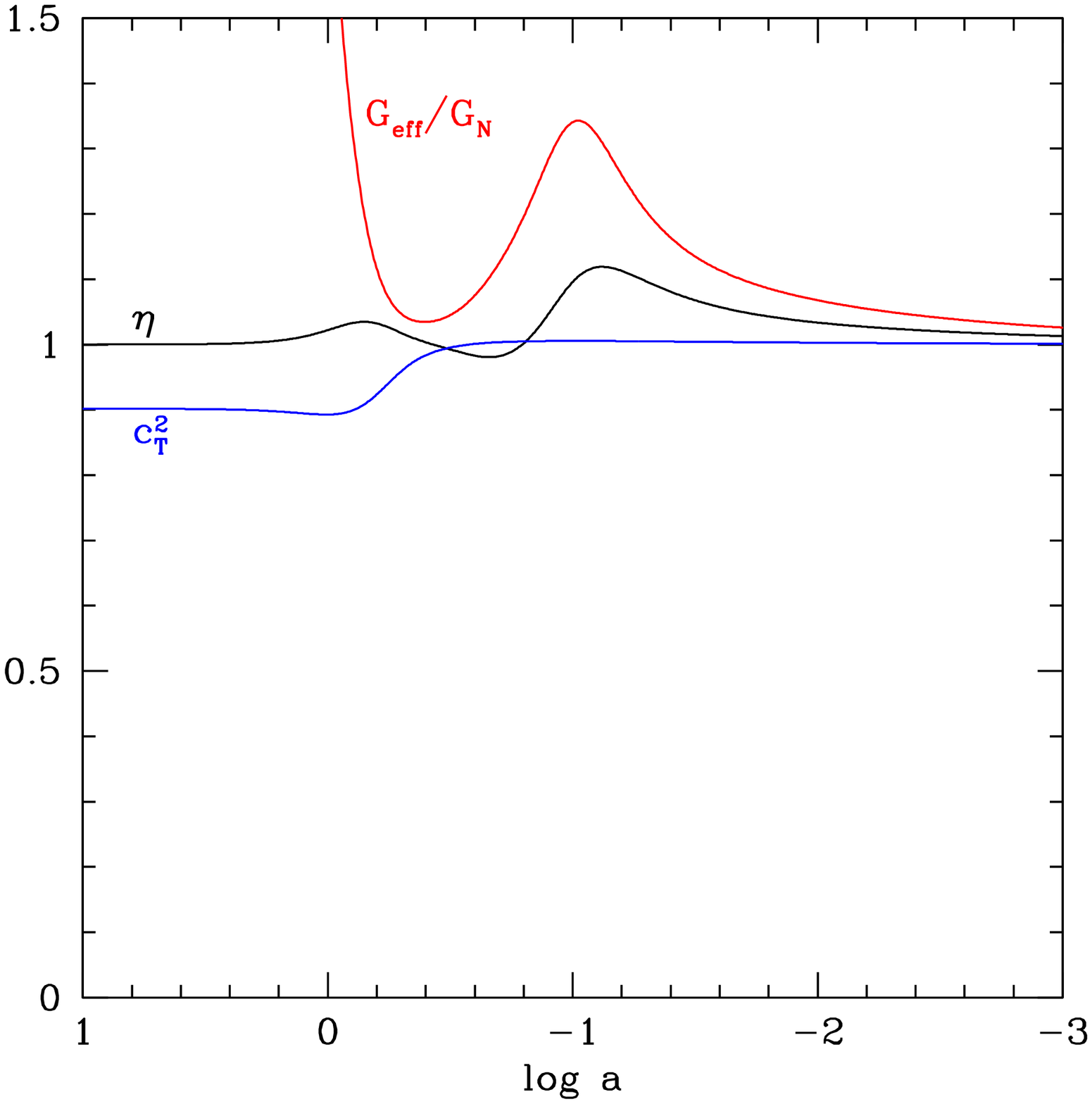} 
\includegraphics[width=0.48\columnwidth]{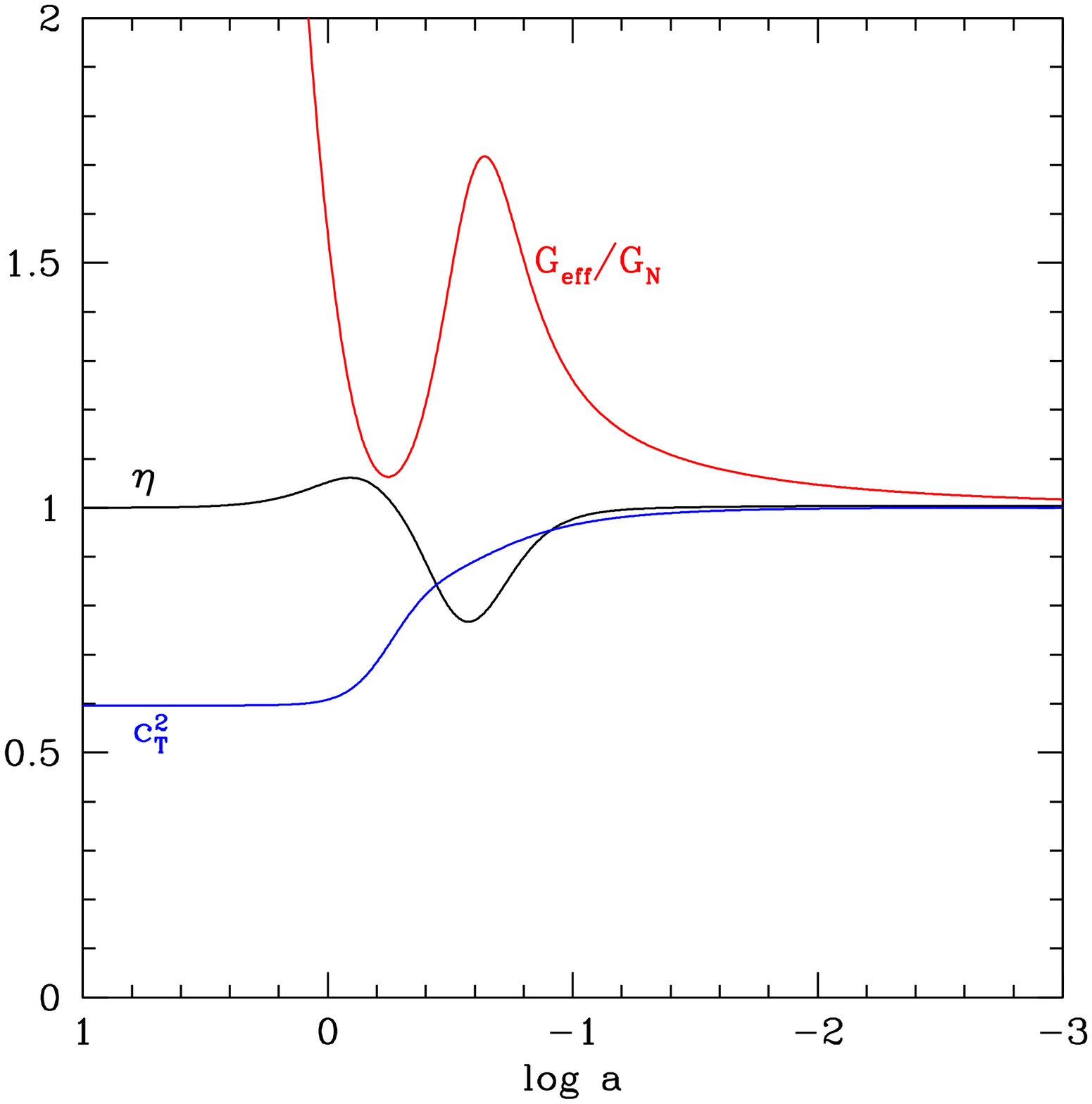}\\ 
\includegraphics[width=0.48\columnwidth]{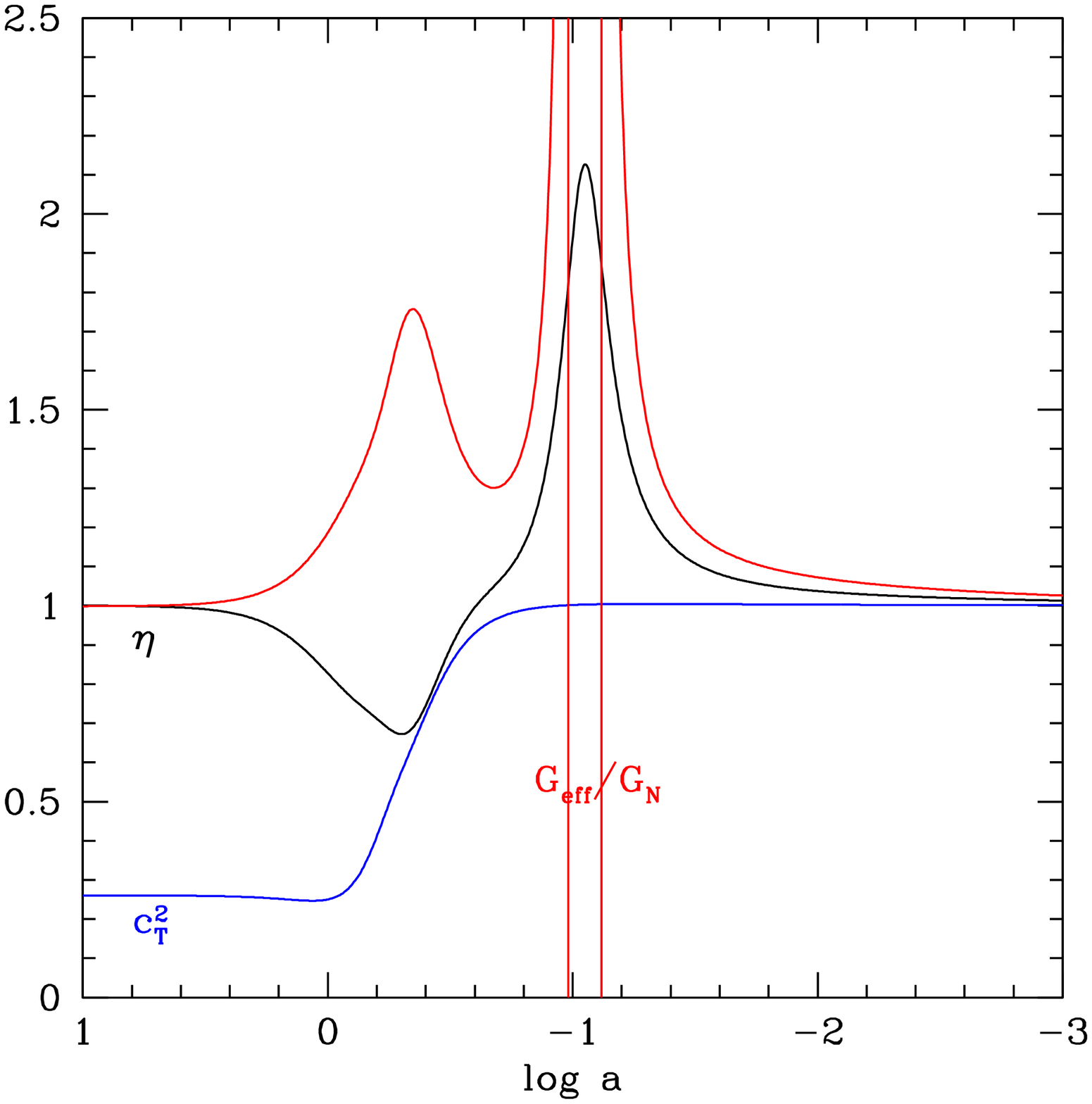} 
\caption{
The observation impacting functions $\geffh$, $\eta$, and $c_T^2$ are 
plotted vs the logarithm of the 
scale factor for the Galileon gravity cases corresponding to Fig.~1, 2, 
and 3 respectively of \cite{slip}. These functions can have nonmonotonic 
dependence and cannot be fit accurately by power law or few parameter 
forms. The values of $\geffh$ asymptote to a 
constant dS value outside the box in the first two panels. 
} 
\label{fig:ctetag} 
\end{figure}

The observationally related functions themselves are even further from 
a power law or simple form. Figure~\ref{fig:ctetag} shows the key 
functions of the gravitational coupling strength $\geffh$, gravitational 
slip $\eta$, and tensor wave speed squared $c_T^2$ that directly affect 
observables such as growth and gravitational wave propagation (including 
cosmic microwave background B-mode polarization). (This figure puts in 
a single view some results shown in \cite{slip} and \cite{gal1112}.) 
Appendix~\ref{sec:apxgalct} discusses the implications of $c_T<1$.

\subsection{Early time limit} \label{sec:early} 

Without knowing the full equations of motion of the specific theory 
we cannot make definite statements about the early time limit, i.e.\ in 
the matter dominated era before cosmic acceleration takes hold. (In the 
previous section we did do this by specializing to the covariant Galileon 
case.) However, we give the following plausibility argument. 

Let us assume that in the early time limit all beyond Einstein-Hilbert 
factors in the action that are nonzero are of the same order of magnitude 
(as one might expect in a natural theory without hierarchy issues, but see 
our further justification below based on the Galileon case). Specifically, 
this means their characteristic scales are the same (if not zero): 
\be 
\Lambda \sim c \sim M_2^4 \sim H\bar M_1^3 \sim H^2 \bar M_2^2 \sim 
H^2 \bar M_3^2 \sim H^2\hat M^2 \sim H^2 m_2^2 \sim H^2 (m_0^2\Omega-m_p^2)\ ,
\label{eq:equalmag} 
\ee 
this makes the not-unreasonable assumption that the early time evolution of 
the modifications is driven by the Hubble expansion time scale. 

The background, Friedmann Eq.~\eqref{eq:friedh} implies that 
$c/(H^2 m_p^2)\sim\oode$, where $\Omega_{\rm de}$ is the fractional 
contribution of the effective dark energy density to the total energy 
density. This allows us to translate any of the mass terms into an order 
of magnitude in terms of $\ode$. 

Furthermore, as the dominant 
time scale in the {\em matter dominated epoch\/} is the Hubble time, we assume 
that time derivatives give rise to a factor $H$, in particular that for the 
kinetic term $\dot X\sim HX$, i.e.\ $\ddot\phi\sim H\dot\phi$, where $H \sim 1/t$ 
during matter domination. 
The importance of this scaling is well known from quintessence \cite{paths} and 
is in contrast to the situation in the EFT of Inflation where the inflaton is 
assumed to dominate the expansion at all times. 

We then see that the other fractional background contribution to the Friedmann 
expansion Eq.~\eqref{eq:friedh} is also 
\be 
\frac{\dot\Omega}{H\Omega}\sim \frac{H(m_0^2\Omega-m_p^2)}{Hm_p^2} 
\sim\oode \ . 
\ee 

Let us turn to the observable functions. From Eq.~\eqref{eq:ctgen} we 
readily see that in the early time limit the tensor wave speed 
\be 
c_T^2\approx 1-\frac{\bar M_3^2}{m_p^2}\sim 1+\oode \ . 
\ee 
For the gravitational slip, evaluating the orders of magnitude of the 
$A_i$, $B_i$, and $C_i$ from Appendix~\ref{sec:apxabc}, 
including our time derivative prescription, gives 
\be 
\bar\eta\sim \frac{m_p^2 H^2\ode +m_p^2 H^2\ode^2}{m_p^2H^2\ode+m_p^2 H^2\ode^2} 
\sim 1+\oode \ . 
\ee 
In particular this can be verified quickly for the background only case of 
Eq.~\eqref{eq:etabgd}. Since $(\dot\Omega/\Omega)\sim H\ode$ then 
$(\dot\Omega/\Omega)^2\sim H^2\ode^2\ll c/(m_0^2\Omega)\sim H^2\ode$ and 
thus 
\be 
\bar\eta_{\rm bg}\approx 1+\frac{(\dot\Omega/\Omega)^2}{c/(m_0^2\Omega)}\approx 
1+\oode \ . 
\ee 
The background only case of Eq.~\eqref{eq:geffbgd} for the gravitational 
coupling is similar: 
\be 
4\pi G_{\rm eff,bg}\approx \frac{1}{2m_p^2}\left[1+\oode\right] \ . 
\ee 
The general case is 
\be 
4\pi\geff\sim \frac{m_p^2 H^2\ode +m_p^2 H^2\ode^2}{m_p^4H^2\ode+m_p^4 H^2\ode^2}
\sim \frac{1}{2m_p^2}\left[1+\oode\right] \ . 
\ee 

Let us now reexamine our input assumptions. First consider the ``equal 
magnitude'' assumption of the action contributions. Approaching this from 
the converse direction, starting from Eqs.~(\ref{eq:nalpha}) -- (\ref{eq:malpha}), 
one sees that if the property functions $\alpha_i$ are $\oode$ then the 
various action functions (mass terms) do indeed have the relations of 
Eq.~\eqref{eq:equalmag}. To explore further, consider the Galileon case. 
The deviation of the observable functions from the GR values 
going as $\oode$ in the early time limit was derived in \cite{gal1112}. This was 
then extended to the property functions deviating as $\oode$ in 
\cite{slip}. The key point, emphasized by \cite{gal1112}, was that at 
early times {\it one\/} of the Galileon $c_i$ terms was dominant over all 
the others, due to their scaling with $H$ (and $X$). This then led to the 
$\alpha_i$ being dominated by one $c_i$ contribution and at the same time 
the fractional effective dark energy density $\ode$ was proportional to the 
same term. 

Generalizing this from Galileon to Horndeski, we ask whether this means 
that one Lagrangian term ${\mathcal L}_i$, involving the $G_i(\phi,X)$ 
and their 
derivatives, is dominant. In the Galileon case this works since 
each $G_i$ is proportional to a constant $c_i$ times a power law function 
of $X$. For example, $G_5=3c_5 X^2$ and so $X\,G_{5,X}\sim G_5$ and 
$\dot G_5\sim H\,G_5$ since $\dot X\sim HX$. However, the property functions 
$\alpha_i$ and the action functions are mixtures of the ${\mathcal L}_i$. 
(See \cite{bellini} and \cite{13046712} for explicit expressions.) 
One dominant $c_i$ term, and hence ${\mathcal L}_i$, feeds into multiple 
$\alpha_i$ and $M_i$. In particular, the $c_4$ and $c_5$ contributions 
feed into all of them (and we expect the $c_5$ term to dominate at early 
times due to its scaling with the highest power of the expansion rate $H$). 
Thus we have the remarkable result that 
\be 
\mbox{\it The\ dominance\ of\ one\ $c_i$ term implies equal orders of magnitude 
for the action functions.} \notag 
\ee 
From this follows, as we have shown, the deviation of the observable 
functions as $\ode$ at early times. 

In general Horndeski, we have functions $G_i(\phi,X)$ not restricted to 
a constant times a power law in $X$. Still, since the property functions 
and actions functions are all functions of $G_{i,\phi}$ and $G_{i,X}$ then 
whichever $G_{i,\theta}$ term dominates should enter and dominate for all 
of them 
and we expect equal orders of magnitude (or zero). (The one exception is 
$m_0^2\Omega$ which alone depends on $G_4$ without derivatives; but a shift 
in the constant part of $G_4$ is simply a redefinition of the Planck mass.) 
So everything depends on the evolution of the scalar field $\phi$, and 
the only timescale in its equation of motion is the background expansion $H$. 

This indicates that indeed in the early time limit of matter domination, 
the action, property, and observational functions should deviate from their 
GR values proportionally to the fractional dark energy density 
contribution $\ode$. We can restate the above aphorism more generally as 
\be 
\mbox{\it The natural dominance of one $G_i$ implies that in the early time limit, 
\{deviations from GR\} $\propto\ode$\ .} \notag 
\ee 
This raises an interesting question as to whether the $G_i$ should be regarded as 
more elemental constituents of the action than the $M_i$, which are combinations 
of derivatives of them. We leave this for future work. 

We expect exceptions to this early time behavior when: 
\begin{enumerate} 
\item Fine tuning of the theory parameters makes multiple elements 
(e.g.\ $G_i$) comparable to each other, 
\item $H$ evolves such that some subdominant term is no longer 
negligible compared to the dominant term (typically when $H/H_0\gg1$ starts 
to fail, if the mass scales of the $G_i$ are normalized so the constant 
coefficients of the $G_i$ magnitudes are all of order unity), or 
\item Another time scale enters the physics, such as from nonlinear collapse 
or couplings to the matter sector. 
\end{enumerate} 
Note that case 2 occurs even for simple, ``natural'' theories, and 
the condition $H/H_0\gg1$ 
breaks well before the present ($z\gtrsim10$) while still $\ode\ll1$. 
For example, in $\Lambda$CDM $H(z=10)/H_0=20$ while $\ode(z=10)=0.002$. Thus 
using $\alpha_i(t)\propto\ode(t)$ or $(\bar\eta-1)\propto \ode(t)$ in the 
late epoch where most observational data exists is not a valid approximation.

\subsection{de Sitter limit} \label{sec:ds} 

In the opposite, late-time limit the cosmic acceleration should lead to a nearly
dS state. In this 
limit, $\alpha_M=0=\alpha_B'$. From Eq.~\eqref{eq:etafull}
this implies that one necessarily has $\bar\eta=1$. However, we 
see from Eq.~\eqref{eq:geff} 
that this does not imply restoration to GR. Not only may 
$c_T^2-1=\alpha_T$ be nonzero but $\geffh\ne G_N$ is possible. 

In particular, from Eq.~\eqref{eq:geff}
\be 
\frac{G_{\rm eff}}{G_N}-1\rightarrow 
\frac{\alpha_B-2[1-(m_p^2/M_{\star,{\rm dS}}^2)]}{2-\alpha_B} 
\ . \label{eq:gab0} 
\ee 
(We remove the superscript $\Phi$ since $\bar\eta=1$ implies that 
$\geffh=\geffs$.) 
We see that several sources of modification of the Poisson equation and 
growth can occur, from braiding and the asymptotic Planck mass. 

Thus, although gravitational slip is guaranteed to vanish in the dS 
limit, deviations from GR can still occur in the strength 
of gravitational coupling and in the tensor sector. 

In the dS limit, for the Galileon case we can derive 
\bea 
\alpha_{M,{\rm dS}}&=&0 \ , \\ 
\alpha_{B,{\rm dS}}&=&2\,\frac{G(2+A)-1}{G(2+A)+1} \ , \\ 
\alpha_{K,{\rm dS}}&=&6\,\frac{G(2+5A)-1}{G(2+A)+1} \ , \\ 
\alpha_{T,{\rm dS}}&=&\frac{G[6+C-(8/3)A]-1}{G(6-4A)+3} \ , \label{eq:galat} 
\eea 
where $G=G_{\rm eff,dS}/G_N$, $C=(c_2 x^2)_{\rm dS}$, and 
$A=(c_G \bar H^2 x^2)_{\rm dS}$, where $c_G$ allows for derivative coupling. 
We can also write 
\bea 
\frac{M_{\star,{\rm dS}}^2}{m_p^2}&=&\frac{2G+1}{2G}\\ 
\frac{m_0^2\Omega}{m_p^2}&=&\frac{G(2+A)+1}{G(2-(4/3)A)+1} 
\frac{G(12+C-(20/3)A)+2}{6G} \ . 
\eea 
In the uncoupled case, when $c_G=0$, 
note that $\alpha_{B,{\rm dS}}$ and $\alpha_{K,{\rm dS}}$ (and $M_\star$) 
are functions 
of $G$ alone. That is, the dS value of the gravitational coupling 
strength determines these parameters. The tensor parameter $\alpha_T$ has 
further freedom however. 

We are free, for example, to choose $G=1$ (as in the third of our triplets 
of plots) and this implies $\alpha_{B,{\rm dS}}=2/3$ and 
$\alpha_{K,{\rm dS}}=2$ for the uncoupled case. Earlier we had noted that 
$\bar\eta=1$ in the dS limit does not imply a restoration to general 
relativity; here we see that even imposing $G=1$ does not do so. We still 
have $c_T\ne1$ and indeed the Poisson equation for growth is still modified 
due to the presence of the $\alpha_i$ (which can be thought of as braiding 
the scalar and tensor sectors, and giving effective dark energy clustering). 
\\

In summary, the EFTDE makes clear the following general results: 
\begin{enumerate} 
\item In the early time limit, the EFTDE action, property, or observational 
functions deviate from their GR values proportionally to 
$\Omega_{\rm de}$, but only under certain conditions. 
\item In the dS limit, the gravitational slip restores to its 
GR value $\bar\eta=\eta=1$ for Horndeski theories. 
\item In the regime of most observations, in between these limits, there 
is no simple, accurate parametrization of the time dependence of the 
functions. Observations must be confronted theory by theory. 
\end{enumerate}

\section{Goldstones, Decoupling, and the de Sitter Limit} \label{sec:dS} 

We now examine further aspects of EFT. In particular, 
we demonstrate that the Goldstone approach to the EFTDE can explain the important result of Eq.~\eqref{eq25}
that the gravitational slip restores to general relativity in the Horndeski and dS limit, 
and how this result can be altered in more general cases. 
We also discuss how the decoupling limit (or lack of it) is useful for grouping models into universality classes 
and allows one to rule out entire subclasses of models within the EFTDE framework.

\subsection{The Goldstone Approach and Decoupling} 

One advantage of the EFTDE is that introducing the Goldstone boson
associated with the spontaneous breaking of time diffeomorphism invariance by the background evolution  
establishes relationships between operators of the low-energy theory and helps classify models in a universal way.
In addition, often it is possible to take a decoupling limit where the Goldstone scalar becomes the most important degree of freedom 
and the other gravitational (gauge) degrees of freedom can be neglected.  These decoupling limits also help to distinguish different classes of models.

We presented the action in unitary gauge in Section \ref{sec:action}, where we wrote the most general theory for the fluctuations about 
a cosmological background realizing $\Lambda$CDM.  The differences in classes of dark energy theories are then encoded in 
the coefficients of the fluctuations\,\footnote{The one exception is again the parameter $m_0^2 \Omega$, however in moving to the Einstein frame this would indeed correspond to coefficients of perturbations in the matter sector (i.e. the parameters determining the interaction strength and masses of particles).} given in Eq.~\eqref{params}.
The construction of the Goldstone action from the unitary gauge action Eq.~\eqref{eftde} was performed in  \cite{Gubitosi:2012hu,bloomfield}.
Physically, introducing the Goldstone allows us to capture the symmetry breaking resulting from the expansion of the universe.
That is, for observations at small distances / high energy (UV) 
the cosmic expansion of the background is negligible (and so invariant under time-translations\,\footnote{The reader may be confused by the use of the term Goldstone boson, since gravity is a gauge theory.  However, in the decoupling limit the 
gravitational (gauge) fields decouple from the would-be Goldstone resulting in a non-linear sigma model and making this language appropriate. Moreover, in this limit time diffeomorphism invariance reduces to time translation invariance. This is analogous to the situation for electroweak symmetry breaking where similar methods can be used to prove the Goldstone equivalence theorem \cite{Weinberg:1996kr}.}), whereas at large distances / low energy (IR) this symmetry is no longer realized -- this is spontaneous symmetry breaking.  Introducing the Goldstone gives us a way to keep track of the symmetry breaking, the fluid or field responsible for it, and if a decoupling limit exists it is often easier to study this scalar than the full gravitational theory.

To introduce the Goldstone boson we perform the broken time diffeomorphism $t\rightarrow t+\xi^0(t,\vec{x})$ on the action Eq.~\eqref{eftde}.
Because the cosmological background depends on time the parameter $\xi^0$ will appear explicitly in the action for the perturbations. 
We then replace $\xi^0\rightarrow \pi(t,\vec{x})$ everywhere it appears in the action and require that the Goldstone transforms as $\pi \rightarrow \pi - \xi^0$ under diffeomorphisms.
Under the transformation we have
\bea
g^{00}&\rightarrow&g^{00}+2g^{0\mu}\partial_\mu \pi + g^{\mu \nu} \partial_\mu \pi \partial_\nu \pi \nonumber \\
g^{0i}&\rightarrow&g^{0i}+g^{\mu i} \partial_\mu \pi \nonumber \\
\delta K_{ij}&\rightarrow&\delta K_{ij}-\dot{H}\pi h_{ij} - \partial_i \partial_j \pi \nonumber \\
\delta K&\rightarrow&\delta K - 3 \dot{H} \pi -a^{-2} \nabla^2 \pi \nonumber \\
\nabla_\mu &\rightarrow& \nabla_\mu, \; g_{\mu \nu} \rightarrow g_{\mu \nu}, \; R_{\mu \nu \lambda \sigma} \rightarrow R_{\mu \nu \lambda \sigma}, 
\eea
where we have expanded to linear order in $\pi$, and we see that covariant quantities remain invariant.

The resulting action appears in \cite{Gubitosi:2012hu,bloomfield}, but for simplicity let us consider the simple case 
$$M_2(t)=\bar M_1(t)= \bar M_2(t)=\bar M_3(t)= \hat M(t)=m_2(t) =0$$ with the resulting Goldstone action
\be \label{pi_action}
S_\pi=\int d^4x \left[ \frac{1}{2} m_0^2 \Omega(t+\pi) R + \Lambda(t+\pi) - c(t+\pi) \left( \delta g^{00} -2\dot{\pi} + 2\dot{\pi} \delta g^{00} +2 \nabla_i \pi g^{0i} - \dot{\pi}^2 + a^{-2} \nabla_i \pi
\nabla^i \pi + \ldots \right) \right]+S_m,
\ee
where dots indicate terms with higher powers of $\pi$ and $S_m$ is the matter action.
We see that this action is invariant under time diffeomorphisms if we require the Goldstone to transform as $\pi \rightarrow \pi - \xi^0({t,\vec{x}})$, i.e.\ the symmetry is non-linearly realized \cite{Senatore:2010wk}. 
Requiring the symmetry be realized in the UV has forced relationships between the various operators (all the terms in parentheses must have coefficient $c$).  In fact, for the case $\Omega=1$ 
all of the operators are fixed by the background evolution $c$ and $\Lambda$ and there are no free parameters -- recall we must use the equations of motion Eqs.~\eqref{eq:friedh} and \eqref{eq:friedh2} to eliminate $\Lambda$ and $c$ from the action and fix 
them in terms of the $\Lambda$CDM history ($H$ and $\dot{H}$).  In the general case we find
\bea \label{c_eqn}
c(t)&=&-m_0^2 \Omega \left( \dot{H}+\frac{\ddot{\Omega}}{2\Omega} + \frac{ \dot{\Omega}}{2\Omega}H \right)-\frac{1}{2}\rho_m, \\
\label{lambda_eqn}
\Lambda(t)&=&m_0^2 \Omega \left( \dot{H} + 3H^2 + \frac{\ddot{\Omega}}{2 \Omega} + \frac{7 \,  \dot{\Omega}}{2 \, \Omega} H\right) -\frac{1}{2}\rho_m,
\eea
which we see evolve as we pass from matter domination to dark energy domination, and as $\Omega$ evolves.

We also note that in general the Goldstone appearing in the coefficients $\Lambda$ and $c$ in Eq.~\eqref{pi_action} would lead to additional terms.  During dark energy domination their time dependence is negligible, but in the matter dominated phase and during the transition to dark energy this time dependence could be 
important\,\footnote{Depending on what scales and observations we are interested in, oscillations in the dark energy could also lead to important corrections; cf.~\cite{Ozsoy:2015rna} and references within.}.

One of the advantages of the Goldstone approach comes from the decoupling limit.  Deep in the dark energy epoch ($\rho_m \ll H^2 m_p^2$) and taking $\Omega=1$, the leading mixing with gravity comes from an operator
\be
{\hat{\cal{O}}_{\rm mix}}\sim   m_p^2 \dot{H} \dot{\pi} \delta g^{00} 
\sim \dot{H}^{1/2} \dot{\pi}_c \delta g^{00}_c
\ee
as in the case of the EFT of inflation. In the last term we introduced the canonically normalized fields $\pi_c \sim m_p \dot{H}^{1/2} \pi$ and $\delta g^{00}_c \sim m_p \delta g^{00}$.
This implies that if we are interested in scales with energy $E \gg E_{\rm mix} \equiv \dot{H}^{1/2}$ we can neglect the gravitational degrees of freedom and focus entirely on the action for the Goldstone.
For example, this approach was utilized to study the stability of quintessence in \cite{Creminelli:2008wc}, where the authors constructed the EFT of Quintessence by introducing the additional operator $M_2$.  The additional operator implies a shift in the normalization of the field as we introduce the Goldstone \cite{Gubitosi:2012hu}
\be
-c(t) \delta g^{00} + \frac{1}{2} M_2^4 (\delta g^{00})^2 \rightarrow (c(t)+2M_2^4) \dot{\pi}^2 - c(t)(\vec{\nabla} \pi)^2 - 2 (c(t)+2M_2^4) \dot{\pi} \delta g^{00}\ , 
\ee 
leading to a sound speed $c_s^2=c/(c+2M_2^4)$ and the mixing energy becomes
$E_{\rm{mix}}\sim (c(t)+M_2^4)/[(c(t)+2M_2^4)^{1/2} m_p]$.  In the general case 
where $c(t)$ is given by Eq.~\eqref{c_eqn} the behavior of $E_{\rm mix}$ could 
become quite involved, similar to the complexity found in previous sections. 

Introducing the operator corresponding to $M_2$ along with a general time dependence for $\Omega(t)$ leads to the most general EFT of a scalar-tensor theory allowed (including theories like $F(R)$ gravity). However, these EFTs are not very 
interesting for a number of reasons \cite{nima}. In particular, requiring consistency with the equivalence principle and fifth force / solar system constraints places strong restrictions on the free parameters. This typically implies that either 
no new observables (predictions) will result, or the models are already ruled out. However, one lesson is that given the EFTDE approach we see that 
all models of a particular class can be scrutinized by data, in this case characterized by $\Omega(t)$ and $M_2$. These constraints can be further enhanced by also accounting for a correct fit of the background to $\Lambda$CDM as was done in \cite{Mueller:2012kb} to rule out an entire class of Gauss-Bonnet type models.  Moreover, these same parameters enter theoretical considerations like the allowed values of the equation of state and the connection to stability as studied in \cite{Creminelli:2008wc}, and the decoupling limit can make such an analysis more tractable.

\subsection{de Sitter Limit} 

During dark energy domination we can take the dS limit corresponding to $\dot{H}=0$ and $\rho_m \ll H^2 m_p^2$ and the tadpole constraints Eq.~\eqref{c_eqn} and Eq.~\eqref{lambda_eqn} simplify to
\bea \label{c1_eqn}
c(t)&=&-m_0^2 \Omega \left( \frac{\ddot{\Omega}}{2\Omega} + \frac{ \dot{\Omega}}{2\Omega}H \right), \\
\label{lambda2_eqn}
\Lambda(t)&=&m_0^2 \Omega \left(  3H^2 + \frac{\ddot{\Omega}}{2 \Omega} + \frac{7 \,  \dot{\Omega}}{2 \, \Omega} H\right),
\eea
which for $\Omega=1$ would reduce to $c=0$ and $\Lambda=3H^2 m_p^2$.  From above we had the sound speed of the Goldstone $c_s^2=c/(c+2M_2^4)$, which we see also goes to zero (even in the presence of the $M_2$ correction). This demonstrates that there are no propagating scalar degrees of freedom in the dS limit.  

However, 
this conclusion can change if we consider additional corrections from the EFTDE.
Consider the additional operators
\be \label{k_correct}
\int d^4x \sqrt{-g} \left( \frac{\bar{M}_2^2}{2} (\delta K)^2 + \frac{\bar{M}_3^2}{2} \delta K^\mu_{\; \; \nu} K^\nu_{\; \; \mu}    \right),
\ee
with the second term being responsible for the altered tensor wave speed we saw in Section \ref{sec:trans}.
Introducing the Goldstone leads to a term in the action \cite{Cheung:2007st}
\be 
\int d^4x \sqrt{-g} \left(  \frac{\bar{M}_2^2+\bar{M}_3^2}{2}  a^{-4} \left(\partial_i^2 \pi \right)^2 \right). \label{golddpi} 
\ee
This implies that in the dS limit the scalar mode will still propagate even though $c$ vanishes.  
This is because the higher derivatives in the equation of motion correct the dispersion relation at ${\cal O}(k^4)$, yet the EFTDE is perfectly consistent as discussed in 
Appendix \ref{apx:beyond}.  Such terms were first utilized in theories of Ghost Condensation \cite{ArkaniHamed:2003uy}.
Now consider the Horndeski theory where we have $\bar{M}_2^2=-\bar{M}_3^2$ and so the correction term in Eq.~\eqref{golddpi} vanishes.
That is, there is again no propagating scalar degree of freedom.  This explains why in the Horndeski dS limit we saw 
that the slip reduced to the GR result. From this Goldstone approach, we can state: 
\\

 {\em Without higher derivative corrections, scalar degrees of freedom will not propagate in pure dS when $c=0$.}
\\

Alternatively, if we allow $\Omega$ to evolve in time, then from Eq.~\eqref{c1_eqn} we see that $c$ will not vanish in the dS limit. 
However, in this case there is not a healthy decoupling limit of the theory.  Recall that for $\Omega=1$ we have the mixing energy $E_{\rm mix}=\dot{H}^{1/2}$ and so in the dS limit the Goldstone 
decouples from gravity.  However, for $\Omega=\Omega(t)$ (or $\alpha_M\ne0$) we saw above that $E_{\rm{mix}}\sim (c(t)+M_2^4)/[(c(t)+2M_2^4)^{1/2} m_p]$ and using Eq.~\eqref{c1_eqn} we see that the mixing will rely on the time dependence of $\Omega$.  In practice, this means that we must either tune the variation of $\Omega$ to be small (to the point where no modification results) or it will be ruled out by experiment.  In addition to this time dependence we note
that even in the absence of such coupling ($\dot\Omega\to0$ or 
$\alpha_M\to0$) the dS state does not mean that all time derivatives 
can be neglected. For example, in the Galileon case the field kinetic 
term $X=(1/2)\dot\phi^2$ goes to a nonzero constant and so $c_s$ doesn't vanish 
(Eq.~39 of 
\cite{gal1112} gives the cubic equation for the uncoupled Galileon case); 
recall that in Horndeski theory many Lagrangian terms involve $X$ or 
functions of $X$ and these do not vanish in the dS limit. 
Figure~\ref{fig:kin} shows this numerically. 

Finally, even though in the pure dS limit the scalar will not propagate, this does not mean that corrections resulting from Eq.~\eqref{k_correct} won't alter the tensor 
sector. 
Indeed we saw in Eq.~\eqref{eq:ctgen} that the tensor speed depends on $\bar{M}_3$ and so the correction in Eq.~\eqref{k_correct} can alter $c_T$ while maintaining $\bar{\eta}=1$ in the dS limit.
We note that this correction survives the dS limit and the Horndeski limit,
\be
c_T^2 =  \left({1-\frac{\bar{M}_3^2}{m_p^2}}\right)^{-1}.
\ee
However, depending on whether the theory has a healthy
decoupling limit this correction may also vanish. For example, in Ghost Condensation we have the
decoupling limit $m^2_p \rightarrow \infty$ with $\bar{M}_3$ fixed and this correction vanishes and
$c_T \rightarrow 1$.

Turning on additional operators in the EFTDE leads to different behaviors, but in taking the decoupling limit two 
interesting classes of models emerge: those that have a healthy decoupling limit, and those that become strongly coupled and require screening mechanisms.  
An example of the latter is given by Table \ref{tab:models} where the class of models that are DGP-like are captured by introducing the additional operator $\delta g^{00} \delta K$ corresponding to $\bar{M}_1$. It is well known that these theories do not have a {\em healthy\/} decoupling limit.  That is, as we try to take the decoupling limit the Goldstone is found to become strongly coupled 
at solar system scales and one then has to explore possible screening mechanisms to see if GR can be recovered in the UV (see e.g.\ \cite{Joyce:2014kja} for a review).  In the EFTDE this corresponds to coefficients like $\bar{M}_1$ growing so large that the validity of the EFT fails and a new EFT would need to be constructed.  At this point the reader may wonder what the usefulness of the EFTDE is then, if model dependent features such as screening mechanisms must be considered. However, there is a familiar analog from particle physics.  In the search to discover the Higgs boson, we used EFT to parameterize the possible mechanisms behind electroweak symmetry breaking.  In particular, we were able to rule out possible models (such as Technicolor) without knowing the UV completion of such theories.  Whereas now that we know a scalar Higgs in consistent with the data, we can try to find the UV extension of the model (SUSY, etc.).  
Here we have a similar situation.  In particular, it is well known that DGP type models are not compatible with structure formation \cite{Sawicki:2007tf} when accounting for the cosmic acceleration. This implies that any model that falls into the universal class in the EFTDE given by $[\Lambda, c, \Omega, M_2, \bar{M}_1]$ (and with the relations given in the table) is inconsistent with the data.  Thus, as an explanation of cosmic acceleration it is no longer useful to study these models and so their UV completion (screening) is irrelevant. This is the power of the EFTDE. Classes of models can be ruled out in the linear regime and help focus model building.

\section{Conclusions} \label{sec:concl} 

Within the last year or two the theoretical structure of dark energy and 
its relation to observables has been recognized to be far richer than 
previously realized. The tensor sector aspects (gravitational waves, 
e.g.\ CMB B-modes) are 
important and complementary to the scalar sector (matter density and 
lensing). In particular, EFTDE demonstrates that the ``classical'' view 
of dark energy in terms of the expansion history $H(z)$ or the equation 
of state $w(z)$ can be merely $1/5$ -- or less -- of the functional 
information! 

EFTDE provides a well defined, complete framework to deal consistently with a 
whole suite of classes of theories, and without imposing by hand constraints 
such as restricting the number of derivatives. We have related the EFT 
action terms, or functions, to the property function approach and the 
observable, or modified Poisson equation, function approach, providing a 
translation dictionary. This brings together views of theorists, 
phenomenologists, and observers. The expansion history, or $H(z)$ Hubble 
function, can be specified separately. 

At quadratic level in the action, EFTDE has seven functions of time 
besides the Hubble function. Unique observation functions are fewer, 
leading to an inverse problem for how data can constrain theory. We 
need either further theoretical principles (even beyond, e.g., the null 
energy condition or Equivalence Principle) or symmetries, or we must search 
assiduously for new types of observational probes (e.g.\ beyond the linear 
or quasistatic regimes, involving screening on astrophysical scales or 
near horizon scale effects). 

While the restriction to the Horndeski class of theories is not justified 
based on limiting to two derivatives, this class does have a nonperturbative 
formulation that is of interest. Therefore Horndeski theories are worth 
studying. This restricts the EFTDE to four free action functions or 
property functions and a like number of observable functions. However 
another type of inverse problem arises: the data does not have the 
leverage to fit arbitrary functions, and we have shown that there is no 
simple time dependence or low dimensional parametrization expected or in 
practice. This unfortunately obviates much of the literature on constraining 
such theories. We demonstrate that the EFT functions are ratios of sums of 
products of ratios of sums! 

We analyze this impasse in some detail, exploring the reasons in the 
origins of action terms as generically of the same order of magnitude; 
for the Galileon subclass we show how this arises from mixes of more 
elementary components. We evaluate the action, property, and observable 
functions numerically for several Galileon instances to demonstrate 
agreement with the theoretical reasoning, with the conclusion that during 
the redshift range $z=0-10$ where the vast majority of observational 
data arises, the functions cannot be reduced to a few parameters. 

Using EFTDE we can show very general results in two limits: early time 
matter domination and late time de Sitter asymptote. In the early time 
limit we demonstrate that for Horndeski theories the deviations of the 
functions from the GR values will be linearly proportional 
to the effective dark energy density contribution. We also lay out how 
and when this breaks down (and in general it does not hold for beyond 
Horndeski theories). 

For the dS limit we prove that within Horndeski 
theories (but not more generally) the gravitational slip will restore to 
GR, $\eta=1$. But this does not mean the physics is that 
of GR. We give an example where one can even tune the 
gravitational coupling strength to GR, $\geff=G_N$, at 
the same time, so the entire scalar sector appears to be GR, 
but the tensor sector will show deviations. This again highlights the 
complementarity of scalar and tensor observations, e.g.\ galaxy and CMB 
polarization surveys. 

We clarify many of the parameter relationships in both the dS and
Horndeski limits by working within the Goldstone 
approach to the EFTDE. By focusing on this scalar associated with the spontaneous symmetry
breaking by the cosmic expansion we argue that the promising models 
for an alternative to a cosmological constant fall into two groups based
on their decoupling behavior.  In this way the EFTDE places
models into universality classes that can be scrutinized by data. 
We also saw that although models that rely on screening mechanisms require
analysis beyond the level of EFT to be complete, it is valuable to use the EFTDE
to determine whether such models can be consistent with existing data in the linear regime.
A notable example is given by the DGP-like class of models where we know they are inconsistent with structure formation.

In the appendices we provide some additional useful details, including a 
table identifying the presence of action functions for specific theories, 
a simple rationale for why EFT removes the need to count derivatives 
(based on work by Weinberg and others), an explicit summary of the EFT equations of 
motion, what constitutes a ``full'' modification of gravity, and the use of the 
gravi-Cherenkov catastrophe to rule out classes of theories. 

In summary, the effective field theory of dark energy is highly effective 
at helping us think deeply about the nature of cosmic acceleration, and 
derive general results in certain limits, but thus far remarkably ineffective in 
the quest to use data to constrain the origin of cosmic acceleration! 
The field is much richer than appreciated a few scant years ago, and 
even determining $H(z)$ or $w(z)$ exactly is just a waypoint in our 
knowledge. The endeavor for a fundamental understanding will require new 
insights in both theoretical and observational techniques. 

\section*{Acknowledgments}
We thank the organizers of PPC 2015 for hospitality and a productive environment that began this work. EL is supported in part by the U.S.\ Department of Energy, Office of Science, Office of High Energy Physics, under Award DE-SC-0007867 and contract no.\ DE-AC02-05CH11231. GS and SW are supported in part by NASA Astrophysics Theory Grant NNH12ZDA001N, and DOE grant DE-FG02-85ER40237.

\appendix

\section{Specific Theories within EFTDE} \label{sec:operator} 

EFTDE covers many different theories of dark energy and modified gravity. 
Table~\ref{tab:models} reviews the connections in terms of which operators 
from the action enter.

\begin{table}[ht]
  \centering
  \begin{tabular}{|l||c|c|c|c|c|c|c|c|c|}
    \hline
    \textbf{Model parameter} &
    \newline
    $\Omega$ &
    $\Lambda$ &
    $c$ &
    $M_2^4$ &
    $\bar{M}_1^3$ &
    $\bar{M}_2^2$ &
    $\bar{M}_3^2$ &
    $\hat{M}^2$ &
    $m_2^2$
    \\ \hline
    \textbf{Corresponding Operator} &
    \; R \; &
    &
    $\;\; \delta g^{00}$ \;&
    $\;\; (\delta g^{00})^2$ \;&
    $\;\;\delta g^{00} \delta {K}{^\mu_\mu}$ \;&
    $\;\;(\delta {K}{^\mu_\mu})^2$ \;&
    $\;\;\delta {K}{^\mu_\nu} {K}{^\nu_\mu}$ \;&
    $\;\;\delta g^{00} \delta R^{(3)}$\; &
    $\;\;\frac{\tilde{g}^{ij}}{a^2} \partial_i g^{00} \partial_j g^{00}$\;
    \\ \hline
    \hline
    $\Lambda$CDM & 1 & \checkmark & 0 & - & - & - & - & - & - \\ \hline
    Quintessence & 1/\checkmark & \checkmark & \checkmark & - & - & - & - & 
- & - \\ \hline 
    $f(R)$ & \checkmark & \checkmark & 0 & - & - & - & - & - & - \\ \hline 
    $k$-essence & 1/\checkmark & \checkmark & \checkmark & \checkmark & - & 
- & - & - & - \\ \hline 
    Galileon {(Kinetic Braiding)} & 1/\checkmark & 
\checkmark & \checkmark & \checkmark & \checkmark & - & - & - & - \\ \hline 
    DGP & \checkmark & \checkmark $\dagger$ & \checkmark $\dagger$ & 
\checkmark $\dagger$ & \checkmark & - & - & - & - \\ \hline 
    Ghost Condensate & 1/\checkmark & \checkmark & 0 & - & - & \checkmark & 
\checkmark & - & - \\ \hline 
    Horndeski {(Generalized Galileon)} & \checkmark & 
\checkmark & \checkmark & \checkmark & \checkmark & \checkmark $\dagger$ & 
\checkmark $\dagger$ & \checkmark $\dagger$ & - \\ \hline
    Ho\v{r}ava-Lifshitz & 1 & \checkmark & 0 & - & - & \checkmark & - & - & 
\checkmark \\ \hline 
  \end{tabular}
  \caption[List of operators associated with various models]{
    A list of operators required to describe various different models 
(adapted from \cite{bloomfield}). \\ 
    \begin{tabular}{p{2cm}cp{0.7 \textwidth}}
      &\checkmark & Operator is necessary\\
      &- & Operator is not included\\
      &1, 0 & Coefficient is unity or exactly vanishing\\
      &1/\checkmark & Minimally and non-minimally coupled versions of this 
model exist\\
      &$\dagger$ & Coefficients marked with a dagger are linearly related 
to other coefficients in that model by numerical coefficients 
    \end{tabular}
  }
\label{tab:models}
\end{table}

\section{EFT and Validity of Beyond Horndeski Theories} \label{apx:beyond} 

How much should we worry about higher derivative operators within EFT, for 
example those arising in beyond Horndeski theories? The 
arguments below follow those emphasized by \cite{Weinberg:2008hq}. 
The EFT approach is formulated by utilizing symmetries to perform a local expansion of the degrees of freedom in the theory.  The leading term (most relevant operator) is corrected by higher derivative terms, which are suppressed relative to the leading term by powers of the cutoff of the theory $M$.
As an example, consider the EFT of a scalar field with
\be
{\cal L}=-\frac{1}{2}\partial_\mu \varphi \partial^\mu \varphi - \frac{1}{2}m^2 \varphi^2 - \frac{c}{M^2} (\Box \varphi)^2  + \ldots,
\ee
where the missing terms are suppressed by further powers of $M$.
Naively it would seem that the correction term would lead to both higher time and space derivatives.
If the equations of motion contain more derivatives this would seem to require more initial conditions -- suggesting the presence of new solutions, which are often argued to lead to runaway behavior.  

It would also seem that the presence of these terms would lead to the emergence of new degrees of freedom due to the correction term. Indeed, the propagator is
\be \label{prop}
\Pi(k) = \frac{1}{k^2+m^2+\frac{c}{M^2} k^4} \ , 
\ee 
suggesting that in addition to the original scalar of mass $m$ we have a new degree of freedom with mass $\sim M$.  However, such an interpretation is incorrect.  If we restrict our attention to energies $E \ll M$, then the action is a local expansion in powers of $M$, where the dimensionless number $c$ is expected 
to be of order one.  It would then be incorrect to write Eq.~(\ref{prop}) without doing the same expansion, i.e. we should instead write
\be
\Pi(k) = \frac{1}{k^2+m^2}\left[1-\frac{k^2}{M^2}\left( \frac{k^2}{k^2+m^2}\right) + {\cal{O}}\left( \frac{k}{M} \right)^4 \right],
\ee
so there is only a single particle (pole) with mass $m$ which is corrected in powers of $k^2/M^2$.
As discussed in \cite{Weinberg:2008hq} an equivalent way to obtain a consistent set of equations is if we substitute the lower order (derivative) equations of motion into the action to eliminate the higher derivative terms.  These are {\em not\/} special constraints, instead by invoking this method we are self-consistently accounting for the fact that the EFT is a local expansion and 
thus the equations of motion should be as well.  Thus, when working within the EFTDE there is no need to enforce special constraints or relations like those assumed in the Horndeski class of models to avoid instabilities.  For a recent pedagogical introduction and a discussion of the subtleties with time dependence we refer the reader to \cite{Burgess:2014lwa}. 

Although the EFTDE {\em always\/} avoids Ostrogradsky instabilities, one may worry that the restriction $E \ll M$ is a severe limitation.  However, the utility of the approach is it allows possible explanations for dark energy to be matched to observations. The observations most relevant to constraining dark energy lie between the present Hubble scale ($10^{-33}\,$eV) and solar system distances or larger  ($r > 1  \; \rm{AU} \approx 1/(10^{-18} \; \rm{eV}$)) or larger.  Thus, we are interested in experiments probing the EFTDE in the energy range $10^{-33} \; \rm{eV} \lesssim E_{\rm{probe}} \lesssim 10^{-18} \; \rm{eV} $.  For example, in Ghost Condensation (a special case of the EFTDE) if we modify gravity on scales of the solar system or larger $r_c^{-1} \sim 10^{-18} \rm{eV}$
this corresponds to a scale  $M \sim (m_p/r_c)^{1/2} \sim 0.1 \; \mbox{MeV}$ \cite{ArkaniHamed:2003uy}.  
Whereas for 
modifications at the Hubble scale (with $\Lambda=0$) one finds $M\sim (m_p/r_c)^{1/2}=10^{-3} \; \rm{eV}$.  We see these both safely satisfy the requirement $E_{\rm{probe}} \ll M$.  
Thus, for using cosmological observations to probe the EFTDE the requirement $E \ll M$ presents no limitations.  

Another possible concern is whether the EFTDE can account for modified gravity models that rely on a screening mechanism to recover GR at solar system scales.  These effects are captured by the {\em time-dependent\/} evolution of the coefficients Eq.~(\ref{params}).  Indeed, for the specific case of screening in DGP gravity it was shown explicitly in \cite{Gubitosi:2012hu} that the EFTDE captured the same effects. However, it is true that in models where screening effects are present it is necessary to find the UV completion for the EFT. 
The advantage of the EFTDE energy scale in that case is that it allows one to see if the proposed model is valid as a low energy explanation of the cosmic acceleration, and if it passes that test, then 
investigations into its UV completion are made worthwhile as we discuss in Sec.~\ref{sec:dS}.
Finally, we comment that the EFTDE can also capture models where higher derivative operators (corresponding to many of the terms in Eq.~\ref{params}) can become more important than a standard kinetic term.  In this case the scaling dimension of the EFT changes and this can also lead to new models for dark energy -- that nevertheless 
can be completely stable at both the classical and quantum level.

\section{EFT Equations of Motion} \label{sec:apxabc} 

The Newtonian limit must take into account the sound horizon rather than 
simply the Hubble scale, that is, $k \gg aH/c_s$, where the sound speed 
\be 
c_s^2=\frac{c/(m_0^2\Omega)+(3/4)(\dot\Omega/\Omega)^2}{c/(m_0^2\Omega)+(3/4)(\dot\Omega/\Omega)^2+2M_2^4/(m_0^2\Omega)} \ . 
\ee 
Note that the presence of $M_2$ causes $c_s$ to differ from the speed of light 
\cite{bloomfield}; equivalently, only $\alpha_K$ is affected by $M_2$. 

In this limit the equations of motion are 
\bea
  -\frac{k^2}{a^2}\left(A_1 \Phi + A_2  \pi + A_3  \Psi \right)&=&  
{\rho}_m \delta_m, \\
  B_1 \Psi + B_2 \Phi + B_3 \pi &=& 0, \\
  \frac{k^2}{a^2}\left( C_1  \Phi + C_2 \Psi + C_3   \pi \right)&=& 0,
\eea 
where $\delta_m \equiv \delta \rho_m / \rho_m$ and the coefficients are 
given by 
\bea
A_1&=&  2m_0^2\Omega+4\hat{M}^2,\\
A_2&=&  -m_0^2 \dot{\Omega}-\bar{M}_1^3+2H\bar{M}_3^2+4H\hat{M}^2,\\
A_3&=&  -8m_2^2,\\
B_1&=&  -1-\frac{2\hat{M}^2}{m_0^2 \Omega},\\
B_2&=&  1,\\
B_3&=&  -\frac{\dot{\Omega}}{\Omega}+\frac{\bar{M}_3^2}{m_0^2 \Omega} 
\left( H+\frac{2 \dot{\bar{M}}_3}{\bar{M}_3} \right),\\ 
C_1&=&  m_0^2 \dot{\Omega}+2H\hat{M}^2+4\hat{M} \dot{\hat{M}},\\ 
C_2&=&  -\frac{1}{2}m_0^2\dot{\Omega}-\frac{1}{2}\bar{M}_1^3- 
\frac{3}{2}H\bar{M}_2^2-\frac{1}{2}H\bar{M}_3^2+2H\hat{M}^2,\\ 
C_3&=&  c(t)-\frac{1}{2}H\bar{M}_1^3-\frac{3}{2}\bar{M}_1^2\dot{\bar{M}}_1 
+\dot{H}\left( 2\hat{M}^2-3\bar{M}_2^2-\bar{M}_3^2 \right) +2H 
\left( H\hat{M}^2+2\hat{M} \dot{\hat{M}} \right),\nonumber \\ 
&+&\frac{k^2}{2a^2}\left( \bar{M}_2^2+\bar{M}_3^2 \right) \ .  
\eea 

For Horndeski theories we have 
\be 
2\hat{M}^2=\bar{M}_2^2=-\bar{M}_3^2 \quad ;\quad m_2=0 \qquad [{\rm Horndeski}] 
\ee 
and these coefficients simplify significantly.
In particular, the last two coefficients become
\bea
C_2&=&  -\frac{1}{2}m_0^2\dot{\Omega}-\frac{1}{2}\bar{M}_1^3,\\
C_3&=&  c(t)-\frac{1}{2}H\bar{M}_1^3-\frac{3}{2}\bar{M}_1^2\dot{\bar{M}}_1 
+(H^2-\dot{H})\bar{M}_2^2 +2H \bar{M}_2 \dot{\bar{M}}_2\ . 
\eea

In the dS limit of Horndeski theory, where all masses and the Hubble 
parameter freeze to constant values, 
\bea
A_1&=&  2m_0^2\Omega+2\bar{M}_2^2,\nonumber \\
A_2&=&  -\bar{M}_1^3,\nonumber \\
A_3&=&  0, \nonumber \\
B_1&=&  -1-\frac{\bar{M}_2^2}{m_0^2\Omega}, \nonumber \\
B_2&=&  1,\nonumber \\
B_3&=&  -\frac{H \bar{M}_2^2}{m_0^2\Omega}, 
\nonumber \\ 
C_1&=&  H\bar{M}_2^2,\nonumber \\
C_2&=& -\frac{1}{2}\bar{M}_1^3,\nonumber \\
C_3&=&  -\frac{1}{2}H\bar{M}_1^3+H^2 \bar{M}^2_2 \ . \label{horndS}
\eea

\section{Meaning of Modification of Gravity} \label{sec:apxmod} 

Exactly what should be interpreted as a modification of gravity, rather than 
a new scalar field, is not generally agreed upon in the literature. 
For non-trivial $\Omega(t)$ in front of the Ricci scalar, we have a 
scalar-tensor theory.  This does not change the spin-2 nature of the 
graviton, so some authors would argue that it is not a true modification 
of gravity. (We merely point this out; we do not take a particular stand 
on the issue, although we will sometimes call theories that modify 
the tensor sector ``full modifications''.) 
That said, these theories are still of course interesting. 

An example of a full modification comes from turning on the extrinsic 
curvature ${K^\mu}_{\nu}$ terms, which are parameterized by 
$\bar M_2$ and $\bar{M}_3$. 
As \cite{bellini} noted, both types of modification 
lead to parameters that come together in pairs (and so they argued the 
$\alpha$ notation is useful).  But this pairing helps us to see a key 
difference between the modifications.  For example, when we introduce 
the Goldstone boson (scalar) as in Sec.~\ref{sec:dS} we find that there 
is a decoupling limit where if we take 
$m_0 \rightarrow \infty$ while holding $\bar{M}_2$ fixed 
that the scalar will decouple from the rest of gravity.  This allows proposals 
like the Ghost Condensate to get around stability problems (although it has 
other issues). We note this 
ratio always appears in the $\alpha$'s (and this is why we defined our 
parameter $N$).  Whereas, a similar decoupling limit does not exist for 
other theories, such as scalar-tensor theories. In other words, a similar 
limit for $\Omega(t)$ does not exist.  In these cases we will have to 
resort to screening by turning on other operators. 

As an example of why scalar-tensor theories are not full modifications of 
gravity, consider the gravitational wave sound speed from 
Eq.~(\ref{eq:ctgen}). In the absence of the extrinsic curvature terms 
$\bar{M}_3 \rightarrow 0$ we recover $c_T=1$.

\section{Gravi-Cherenkov Issues of Galileons} \label{sec:apxgalct} 

Note that the gravi-Cherenkov limit \cite{nelson,stoica} requiring  
$c_T\gtrsim1$ (really $>-10^{-15}$ from \cite{kimura}) puts strong 
constraints on modified gravity. However, \cite{tscreen} points out that the 
high energy of the radiated graviton involved, $\sim10^{10}$ GeV, may put 
it beyond the scope of the effective field theory. If one does take the 
limit at face value, it appears to rule out the entire class of standard 
(uncoupled) covariant Galileons, as seen below. 

In the de Sitter limit for Galileons (see Eq.~13 of \cite{slip}), 
the tensor speed excess is 
\be 
c_T^2-1=\frac{2E-2A-3F}{1-(3/2)E+A+3F} \ , 
\ee 
where $A$ comes from the derivative coupling term, and $C$, $D$, $E$, $F$ 
are the standard Galileon terms, e.g.\ $F=c_5 \bar H^6 x^5$ -- see 
\cite{gal1112}. Using the equations of motion, in the de Sitter limit we 
have (Eqs.~67-68 of \cite{gal1112}) 
\bea 
3F&=&2D-C-4-2A\\ 
9E&=&8D-3C-10-2A \ , 
\eea 
so we obtain the condition for $c_T^2<1$ to be 
\be 
-2D+3C-16-4A<0 \quad{\rm if}\quad 2D-(3/2)C-4-2A>0 \,. \label{eq:galdsct} 
\ee 
But the second condition, coming from the denominator of $c_T^2$, is 
related to the no-ghost condition when $A=0$ (no derivative coupling, as 
we now assume).  

Having no ghosts requires (Eq.~71 of \cite{gal1112}) 
\be 
2+(3/4)C<D<5+(3/4)C \,. \label{eq:galdsghost}
\ee 
Since $C=c_2x^2<0$ (right sign kinetic term, imposed by positive energy 
density and no ghosts; see the right panel of Fig.~5 in \cite{gal1112}), 
then the lower bound of Eq.~(\ref{eq:galdsghost}) implies $2D>4+(3/2)C$, 
satisfying the second condition of Eq.~(\ref{eq:galdsct}). We are thus 
left with evaluating the first condition of Eq.~(\ref{eq:galdsct}), whether 
$2D>3C-16$. Again we can write the lower bound of the no-ghost condition as 
\be 
2D>(3/2)C+4>(3/2)C+4-[20-(9/2)C]=3C-16 \,, 
\ee 
where the second inequality follows because the quantity in square 
brackets is positive ($C$ being negative). But the final expression is 
exactly the first condition of Eq.~(\ref{eq:galdsct}) and so we see that 
for uncoupled covariant Galileons, $c_T^2<1$. If the gravi-Cherenkov 
condition is valid, this rules out this entire class of modified gravity. 

An alternate, shorter method of reaching this conclusion is to use 
Eq.~(\ref{eq:galat}) to ask whether $\alpha_T<0$. For the uncoupled Galileon 
case this occurs when $G(6+C)<1$. If we insist that the strength of gravity 
be positive, then this requires $C<-6+(1/G)$. From Fig.~5 of 
\cite{gal1112} we always have $C<-6$, and so indeed uncoupled Galileons have 
$\alpha_T<0$ or $c_T^2<1$, leading to a gravi-Cherenkov catastrophe.


\end{document}